\documentclass[aps, superscriptaddress, 12pt, nofootinbib]{revtex4}

\usepackage{graphicx}
\usepackage{amsmath}
\usepackage{psfrag}

\def\slash#1{\setbox0=\hbox{$#1$}               
   \dimen0=\wd0                                 
  \setbox1=\hbox{/} \dimen1=\wd1               
   \ifdim\dimen0>\dimen1                        
      \rlap{\hbox to \dimen0{\hfil/\hfil}}      
      #1                                        
   \else                                        
      \rlap{\hbox to \dimen1{\hfil$#1$\hfil}}   
      /                                         
   \fi}                                         %

\newcommand{\nn}{\nonumber}
\newcommand{\be}{\begin{eqnarray}}
\newcommand{\ee}{\end{eqnarray}}
\newcommand{\bea}{\begin{eqnarray}}
\newcommand{\eea}{\end{eqnarray}}

\begin{document}

\title{Chiral Expansion from Renormalization Group Flow Equations}

\author{Lars Jendges}
\affiliation{Institute for Theoretical Physics, University of
  Heidelberg, Philosophenweg 19, 69120 Heidelberg}
\author {Bertram Klein} 
\affiliation{GSI, Planckstrasse 1, 64291 Darmstadt}
\author{Hans-J\"urgen Pirner}
\affiliation{Institute for Theoretical Physics, University of
  Heidelberg, Philosophenweg 19, 69120 Heidelberg}
\affiliation{Max-Planck-Institut f\"ur Kernphysik, Saupfercheckweg 1,
  69117 Heidelberg}
\author{Kai Schwenzer}
\affiliation{Institute for Physics, University of Graz, Universit\"atsplatz 5, A-8010 Graz} 

\date{\today}

\begin{abstract} 
We explore the influence of the current quark mass on observables
in the low energy regime of hadronic interactions within a
renormalization group analysis of the Nambu--Jona-Lasinio model in its bosonized form.  
We derive current quark mass expansions for the pion decay
constant and the pion mass, and we recover analytically
the universal logarithmic dependence. A numerical
solution of the renormalization group flow equations enables us to
determine effective low energy constants from the model.
We find values consistent with the phenomenological estimates used in
chiral perturbation theory. 
\end{abstract} 

\maketitle

\section{Introduction}
\label{sec:intro}

Chiral Perturbation Theory ($\chi$PT) is an effective field theory
which describes the low energy limit of the physics of the strong interaction where its theoretical description, Quantum Chromodynamics (QCD), is entirely
non-perturbative. It is based on the expansion of an effective Lagrangian
in terms of small external
momenta and the pion  mass. A finite pion mass arises from the small explicit breaking of chiral symmetry. As an
effective field theory, $\chi$PT is defined for scales much smaller
than the chiral symmetry breaking
scale $\Lambda_\chi \sim 1 \; \mathrm{GeV}$, and it
is renormalizable order by order in the small momentum expansion
\cite{Weinberg:1978kz, Gasser:1983yg, Gasser:1984gg}. 
 The physics at low momenta 
is completely dominated by the light pions. 
Other fluctuations are
separated from the light Goldstone bosons by a mass gap and therefore 
suppressed. Due to the constraints of chiral symmetry, all quantum corrections are due to pions and depend on the pion mass parameter.
However, going beyond leading order in the chiral
expansion, the number of undetermined coupling constants in the
chiral Lagrangian increases rapidly.

The chiral expansion in small pion masses and momenta
gives rise to terms that are non-analytic in the pion mass, i.e  logarithmic
corrections appear at the one-loop level.
In leading order, these are universal corrections and entirely determined
by the symmetry. Therefore, they should be reproduced by any theory that
describes the low energy domain within a suitable approximation.
On the lattice, the appearance of these chiral logarithms has not yet
been unambiguously demonstrated, but recent results remain compatible
with the existence of such corrections \cite{Luscher:2005mv}.
It is our purpose to show that the
logarithmic terms are recovered in the pion sector of the
bosonized Nambu--Jona-Lasinio (NJL) model \cite{Nambu:1961fr,
  Nambu:1961tp}, and to determine them in a Renormalization Group (RG)
approach.

An effective field theory can be considered as the limit
of a more fundamental theory at low energy. Only fields relevant in this energy
range are retained, while all other degrees of freedom are integrated
out. The effects of these higher modes enter into low energy couplings (LECs) which
multiply the remaining operators in the low energy theory. 
These LECs have to be obtained either from
experiment \cite{Gasser:1984gg}, or by non-perturbative methods from
the underlying gauge 
dynamics of QCD. Because of the strong gauge coupling at
low energies and the emergence of collective hadronic degrees of
freedom this remains a difficult task.

Recent progress in a non-perturbative description has been made using Dyson-Schwinger Equations (DSE)
\cite{Alkofer:2000wg, Fischer:2002hn, Fischer:2003rp}, or RG flow equations 
\cite{Pawlowski:2003hq, Fischer:2004uk, Gies:2005as, Braun:2005uj, Braun:2006jd}. 
Numerical lattice simulations are limited by the 
computational effort and require extrapolations to large volumes,
small quark masses, and small lattice spacing. The extrapolation to
small quark  mass is actively pursued
\cite{Luscher:2005mv} and an important topic for $\chi$PT.  
Finite volume effects have been addressed in $\chi$PT as well, both for the nucleon
\cite{Hemmert:2002uh} and for the pion 
\cite{Colangelo:2002hy, Colangelo:2003hf, Colangelo:2004xr,
  Colangelo:2005gd}, and in the framework of the quark meson 
model with RG methods \cite{Braun:2004yk, Braun:2005gy, Braun:2005fj}.

Due to the importance of the low energy couplings on one hand, and
the difficulty of obtaining them from QCD on the other, there has been a
strong interest in phenomenological models \cite{Ebert:1985kz,
  Ecker:1988te, Hansson:1990jy, Schuren:1991sc, RuizArriola:1991gc,
  RuizArriola:1991bq, Klevansky:1992qe, Schuren:1993aj,
  Mueller:1994dh, Hippe:1995hu, Bijnens:1995ww, Jungnickel:1997yu, Llanes-Estrada:2003ha}. RG methods are uniquely suited to interpolate
 between the low energy effective
theory and QCD \cite{Bijnens:1995ww}.

There are \emph{bottom-up} or \emph{top-down} models.
In the bottom-up approach, the starting point is $\chi$PT. 
Additional vector,
axial-vector and scalar meson resonances may be
included. In ref.~\cite{Ecker:1988te}, it was argued that the coupling
constants of the chiral Lagrangian in next-to-leading order could be
accounted for mainly by contributions of vector meson resonances. In
\cite{Jungnickel:1997yu} it was found that in a linear meson
model the low energy couplings are dominated by the exchange
of scalar mesons. 
Such descriptions rely entirely on a hadronic
picture. 
On the other hand, the top-down approach starts from models with
quark degrees of freedom, which generate mesonic degrees of freedom by quark dynamics. These approaches are essentially based on the
NJL model \cite{Nambu:1961fr, Nambu:1961tp},
which includes a fermionic self-interaction. Meson fields arise from a
bosonization of the
quark interaction with a Hubbard-Stratonovich transformation
\cite{Hubbard:1959ub}. Pions appear explicitly as the Goldstone bosons of
spontaneous breaking of the chiral symmetry through a quark
condensate. For a review of this class of models, see
e.g. \cite{Bijnens:1995ww, Klevansky:1992qe}. A comparison with $\chi$PT has
been undertaken 
e.g. in \cite{Schuren:1991sc, RuizArriola:1991gc, RuizArriola:1991bq,
  Mueller:1994dh, Hippe:1995hu}. More recently, a third approach based
on a weak form of the ADS/CFT correspondence \cite{Aharony:1999ti} shows a remarkable
agreement with hadron phenomenology \cite{DaRold:2005zs}.

In the RG treatment \cite{Jungnickel:1995fp, Berges:1997eu, Schaefer:2004en,
  Meyer:2001zp} of NJL-based
models, chiral symmetry is dynamically broken, and the pions emerge
naturally in 
the RG flow of the bosonized model. In order to capture the low energy
dynamics dominated by the pions, the meson dynamics has to
be included in the RG flow. This goes beyond the leading order of the
often employed large-$N_c$ approximation \cite{'tHooft:1973jz,
  'tHooft:1974hx, Witten:1979kh}. Only the  
inclusion of such $1/N_c$ corrections \cite{Meyer:2001zp} makes it
possible to derive  the leading-order logarithmic corrections in the
quark mass expansion of low energy observables.  
In particular, our results show
that the wave function renormalization $Z_\phi$ for the derivative term of the mesonic fields is essential, since it enters into the 
expression for the pion decay constant $f_\pi$, the coupling
constant for the lowest-order derivative term in $\chi$PT (compare \cite{Jungnickel:1997yu}).

In a comparison of RG results to $\chi$PT, one would like to
calculate the same effective couplings.
Already the lowest order $\chi$PT Lagrangian contains information about
arbitrary pion $n$-point functions, which makes an order-by-order
comparison with the results of RG methods difficult, since the
effective RG action is expanded in $n$-point functions.
Alternatively, one can proceed in the same way as in the comparison with
experimental results. By calculating observables and identifying the
influence of the low energy constants
one can compare effective values for these couplings. In this way one
does not actually have to project the RG flow onto the operators used
in the chiral expansion. We will employ this method in the present paper.
Our comparison between $\chi$PT and the NJL model will be
limited to the two flavor case and of course to only those observables that
appear in both theories.
In the specific case of the bosonized NJL model, which we define at a
large UV scale, there are several free parameters. The
UV  parameters are  
fixed by requiring that the model should reproduce the physical values
for the pion decay constant $f_\pi$ and the pion mass $m_\pi$, after all quantum fluctuations have been 
integrated out. While the actual values of $f_\pi$ and $m_\pi$ are
input parameters to the model and not predicted, their predicted dependence
on the current quark mass tests whether the model as a low energy description of meson
physics is compatible with other approaches. Since the low energy
constants depend on the underlying
short distance theory, they differentiate whether the model
contains the relevant dynamics for the emergence of the low energy physics.
Such a consistency check can put the model on a more sound footing, even
without a direct connection to the gauge dynamics of QCD.

The main part of this paper is organized as follows:
After a short review of some results of chiral perturbation theory at
one-loop order in section \ref{sec:chPT}, we will present the RG approach
to the bosonized NJL model and introduce the RG flow equations in
section  \ref{sec:RG}.
Analytic results for the non-analytic dependence of the pion mass
and the pion decay constant on the symmetry breaking parameter are
given in section \ref{sec:analytical}.
Numerical results on the same
quantities as function of the current quark mass are given in
section \ref{sec:numerical}. We will demonstrate that the results from
the NJL model
for the couplings that appear in the pion mass and the pion decay
constant are compatible with the values used in $\chi$PT calculations.  
We present a summary and our conclusions in
section \ref{sec:conclusions}. Details of the derivation of the RG
flow equations can be found in the appendix.

\section{Results from Chiral Perturbation Theory}
\label{sec:chPT}

In this section we give a short summary of some principles and
results of $\chi$PT at the one-loop level following
the work of Gasser and Leutwyler \cite{Gasser:1983yg} and the recent more pedagogic article \cite{Leutwyler:2001}. 
Starting point is the Lagrangian $\mathcal{L}_{\chi {\rm PT}}$ of the non-linear sigma model in
the O(4) chiral field $\hat{U}=(\hat{U^0},\hat{U^i})$ of unit length expanded  
up to fourth order in momentum. 
We give this Lagrangian with the 10 phenomenological coupling
constants ($l_i, h_i$) here for later comparison with the RG-results
\begin{align}
&\mathcal{L}_{\chi PT} =\frac{f_{\pi0}^2}{2}\nabla_{\mu} \hat{U^T} \nabla^{\mu}  \hat{U}
+2 B f_{\pi0}^2 \ (s^0  \hat{U^0}+p^i  \hat{U^i})
+ l_1 \  (\nabla^{\mu}  \hat{U^T} \nabla_{\mu}  \hat{U})^2\nonumber\\
&+ l_2 \ (\nabla^{\mu}  \hat{U^T} \nabla^{\nu} \hat{U})(\nabla_{\mu} \hat{U^T}
\nabla_{\nu} \hat{U})+l_3\ (\chi^T  \hat{U})^2+ l_4\ (\nabla^{\mu}
\chi^T\nabla_{\mu}  \hat{U})\nonumber\\
&+l_5\ ( \hat{U^T} F^{\mu \nu} F_{\mu \nu} \hat{U})+l_6\
(\nabla^{\mu} \hat{U^T} F_{\mu \nu} \nabla^{\nu} \hat{U})+l_7\ (\tilde{\chi}^T  \hat{U})^2\nonumber\\
&+h_1 \chi^T
\chi+ h_2 tr F_{\mu \nu}F^{\mu \nu}+ h_3\tilde{\chi}^T\tilde{\chi}\label{Lchpt},
\end{align}
with 
\begin{align}
\nabla_{\mu} \hat{U^0}&=\partial_{\mu} \hat{U^0}+a^i_{\mu} \hat{U^i}\nn\\
\nabla_{\mu} \hat{U^i}&=\partial_{\mu} \hat{U^i}+\varepsilon^{i k l }v_{\mu}^k  \hat{U^l}
-a^i_{\mu} \hat{U^0}\nn\\
\chi &= 2 B(s^0,p^i),\nn\\
\tilde{\chi} &= 2 B(p^0,-s^i)\nn\\
F_{\mu \nu} \hat{U}&=(\nabla_\mu \nabla_\nu -\nabla_\nu \nabla_\mu)\  \hat{U},\nn
\end{align} 
where $v_{\mu}(x),a^i_{\mu}(x),s(x)$ and $p^i(x)$ are external vector,
axialvector, scalar and pseudoscalar fields. Without isospin breaking the constants  $l_7$ and $h_3$ are zero.\\
The constant $f_{\pi0}$ is the pion decay constant in the chiral
limit, which in the literature of chiral perturbation theory usually
is called $F$, and
$B$ parametrizes the chiral condensate in the chiral limit,  
\be
B=-\frac{1}{2}\frac{\left\langle{\bar{q}q}\right\rangle_0}{f_{\pi 0}^2}=-\frac{1}{2}\frac{\left\langle{\bar{u}u+\bar{d}d}\right\rangle_0}{f_{\pi 0}^2}.
\ee
In the two flavor
case treated in this work, we consider the masses of the two lightest
flavors to be equal. In order to take the quark mass term 
into account one expands the scalar external field around $s=m_c$,
where $m_c$ is the average current quark mass:
\begin{eqnarray}
m_c &=& \frac{1}{2} (m_u + m_d)\label{mc}
\end{eqnarray}
One obtains the Gell-Mann--Oakes--Renner relation, where the parameter $M$ is
the pion mass in lowest order in the current quark mass
\begin{equation}
M^2 =2 m_c B =  - m_c \frac{\left\langle{\bar{q}q}\right\rangle_0}{f_{\pi0}^2}.\label{gmor} 
\end{equation}
The non-linear sigma model is not renormalizable, but it is possible to
make the theory finite at every loop order by introducing appropriate counter terms. In chiral
perturbation theory this renormalization is usually done by
dimensional regularization. Up to one-loop order, one
needs the above ten low energy constants $l_i$ and $h_i$ which depend logarithmically
on the renormalization scale $\mu$,

\begin{align}
 l{_i}^r &= {\gamma_i \over 32 \pi^2}(\bar{l}_i+\log{{M^2 \over
    \mu^2}})\ \ \ i=1,... ,7 \label{l}\\
 h{_i}^r &= {\delta_i \over 32 \pi^2}(\bar{h}_i+\log{{M^2 \over
    \mu^2}})\ \ \ i=1,2,3\label{h}.
\end{align}
The low energy constants $\bar{l}_i$ are independent of the renormalization scale and are related to physical observables. The constants $\bar{h}_i$ have no direct physical
relevance, but follow from the renormalization procedure. To go to higher loop
order, an increasingly larger number of constants is needed.
By computing Greens functions from chiral perturbation
theory in the limit of very small external momenta, one derives
physical observables which are expanded  in $M^2$. As shown above,
$M^2$ is related to the current quark mass $m_c$.
For future reference, we quote the chiral expansions for
the pion decay constant, the chiral quark condensate and the pion
mass,

\begin{align}
f_\pi &= f_{\pi0} \left(1-\frac{M^2}{16 \pi^2 f_{\pi0}^2} \log{\frac{M^2}{\mu^2}}+
\frac{l{_4}^r}{f_{\pi0}^2} M^2 +{\cal O}(M^4)\right) 
\label{Fpi}\\
\left\langle{\bar{q}q}\right\rangle &= \left\langle{\bar{q}q}\right\rangle_0
\left(1-\frac{3 M^2}{32 \pi^2 f_{\pi0}^2}
\log{\frac{M^2}{\mu^2}}+\frac{2(h_1^r+l_3^r)}{f_{\pi0}^2}M^2 +{\cal
  O}(M^4)\right)\label{qq} \\
m^2_\pi &= M^2 \left(1+\frac{M^2}{32 \pi^2 f_{\pi0}^2} \log{\frac{M^2}{\mu^2}} +
2 \frac{l{_3}^r}{f_{\pi0}^2} M^2+{\cal O}(M^4)\right)\label{eq:Mpi}.
\end{align}
We will compare our RG results with these expansions.

The expansions are characterized by the so-called chiral logarithms and
by higher order terms in the expansion parameter $M^2$. It is possible to
combine the chiral logarithms with the linear term in $M^2$
(cf. equations (\ref{l}) and
(\ref{h})), so that the constants which appear do not depend on the
renormalization scale $\mu$ any more, but only on $M^2$,
\begin{align}
f_\pi &= f_{\pi0} \left(1+\frac{M^2}{16 \pi^2 f_{\pi0}^2}
  \bar{l_4}+{\cal O}(M^4)\right) 
\label{Fpi2}\\
\left\langle{\bar{q}q}\right\rangle &= \left\langle{\bar{q}q}\right\rangle_0
\left(1+\frac{M^2}{32 \pi^2 f_{\pi0}^2} (4 \bar{h}_1-\bar{l}_3) +{\cal
 O}(M^4)\right)\label{qq2} \\
m^2_\pi &= M^2 \left(1-\frac{M^2}{32 \pi^2 f_{\pi0}^2} \bar{l_3}
+{\cal O}(M^4)\right)\label{eq:Mpi2}.
\end{align}
The renormalization scale-independent constants $\bar{l_i}$ are parametrized in
terms of the constants $\Lambda_i^2$, 
\begin{align}
&\bar{l_i}= \log{\frac{\Lambda_i^2}{M^2}}\label{liconstants}\\ 
&\Lambda^{\scriptscriptstyle {\chi PT}}_3=0.59^{+1.40}_{-0.41} \; \rm{GeV}\; , \;\Lambda^{\scriptscriptstyle {\chi PT}}_4=1.26\pm 0.14 \; \rm{GeV}
\end{align}
The $M^2$-dependence is contained in the logarithms, $\Lambda^{\scriptscriptstyle {\chi PT}}_3$ and
$\Lambda^{\scriptscriptstyle {\chi PT}}_4$ are scale-independent
constants, as given in
\cite{Leutwyler:2001}, while $\bar{h}_1$ is not determined in chiral
perturbation theory since $\left\langle{\bar{q}q}\right\rangle$ is not
a physical quantity.

\section{Renormalization Group flow equations for the bosonized NJL model}
\label{sec:RG}

In this section, we  give a brief overview of the renormalization group approach to the NJL model. The detailed approximation scheme and the derivation of the flow equations are presented in \cite{Schwenzer:2005, Schwenzer:2003diss}.
Here we only review the main results and give the flow equations that are used for the analysis of the chiral low energy expansion in the next section.

The basic idea of the renormalization group is to describe the
dependence of an effective action on an infrared cutoff scale, and to follow
the evolution of the couplings under a change of this 
cutoff scale. In this way, quantum fluctuations are integrated out in
a systematic way, and a theory with bare couplings defined at a given UV
scale is transformed into an effective low energy theory in which all
quantum fluctuations with large momenta are already included in the couplings
\cite{Wetterich:1992yh, Berges:2000ew}.

Our starting point is the NJL model \cite{Nambu:1961fr, Nambu:1961tp} with quark fields
interacting via a point-like four-fermion  interaction.
We use this model of dynamical chiral symmetry breaking to describe the
transition from quark fields to hadronic degrees of freedom for scales
$k  \lesssim 1 \;\mbox{GeV}$.   
The Lagrangian reads in Euclidean spacetime
\be
{\mathcal L}_{\mbox{\tiny NJL}} &=& \bar{q}(\slash{\partial}+m_c)q-g_{\mbox{\tiny NJL}}\left(
(\bar{q}q)^2+(\bar{q}i\vec{\tau}\gamma_5 q)^2\right). \label{eq:lnjl}
\ee   
This Lagrangian contains the dominant chirally symmetric four-fermion
interaction. Chiral symmetry is preserved by the interaction term, but
broken explicitly by a finite current quark mass $m_c$. By
bosonization of the quark fields at the UV scale, we obtain
the associated linear $\sigma$-model \cite{Ripka:1997}, where the
chiral symmetry of the quark fields ($\bar q,  q$) is for two quark
flavors reflected in an O(4)-symmetry
of the meson fields $\Phi=(\sigma, \vec{\pi})$,
\be
{\cal L}^{\mbox{\tiny UV}}= \bar{q} \slash{\partial} q
+g_{\mbox{\tiny UV}}\bar{q}(\sigma+i \vec{\tau}
\vec{\pi}\gamma_5)q+U_{\scriptstyle{\mbox{\tiny UV}}}(\Phi^2,\sigma,k_{\mbox{\tiny UV}})
\label{eq:lqmm}
\ee
with
\be
U_{\scriptstyle{\mbox{\tiny UV}}}(\Phi^2,\sigma,k_{\scriptstyle{\mbox{\tiny UV}}})&=& \frac{m_{\mbox{\tiny UV}}^2}{2}(\sigma^2+\vec{\pi}^2)-\delta
\sigma,\\
\delta&=&\frac{m_{\scriptstyle{\mbox{\tiny UV}}}^2}{g_{\scriptstyle{\mbox{\tiny UV}}}}m_c.
\ee
The parameter $\delta$ depends on the current quark mass and describes
explicit chiral symmetry breaking.

We sketch the derivation of renormalization group flow
equations for the effective action $\Gamma$ of the bosonized NJL model
in the Schwinger proper-time regularization scheme \cite{Liao:1994fp}. 
The effective action is the generating functional of the one-particle
irreducible Greens functions of quarks and meson
fields. The derivative expansion of the effective
action up to next-to-leading order is given by 
\be
\Gamma [\bar{q},q,\Phi]= \int d^4 x \Bigg( U(\Phi^2,\sigma)+ \frac{1}{2}
Z_\Phi^{ab}(\Phi^2) (\partial_\mu \Phi^a)  (\partial_\mu \Phi^b)\nn\\
+G(\Phi^2) & \bar{q}(\sigma+i \vec{\tau}
\vec{\pi}\gamma_5) q+Z_q(\Phi^2) \ \bar{q}
\slash{\partial}q \Bigg).\label{GAMMA}
\ee
In general, the effective wave-function renormalization factors
$Z^{ab}_\Phi(\Phi ^2)$,
$Z_q(\Phi^2)$, as well as the effective Yukawa coupling $G(\Phi^2)$ in
this action, are functions of the bosonic fields $\Phi$ and thereby
include arbitrary many ``elementary'' couplings - in full
analogy to the effective potential $U(\Phi^2,\sigma)$. In
ref.~\cite{Schwenzer:2005} a solution is given for this general
case. In this work we only treat the effective potential
$U(\Phi^2,\sigma)$ as
field-dependent, while we approximate the other couplings by
taking only their first, $\Phi$-independent term into account, i.e.
\begin{equation}
Z_\Phi^{ab}(\Phi ^2)\equiv Z_\Phi(\sigma_0^2) \ \delta^{ab}, \;\;\;
G(\Phi^2)\equiv G(\sigma_0^2),\;\;\;
Z_q(\Phi^2)\equiv Z_q(\sigma_0^2),
\end{equation}
where $\sigma_0$ denotes the vacuum expectation value from the effective
potential $U$. We use the same wavefunction renormalization $Z_\Phi(\sigma_0^2)$ for all mesonic fields.
The flow equations for the bosonized NJL model in the proper-time renormalization group scheme are derived in ref.~\cite{Schwenzer:2005}. These equations for the individual couplings of the effective action take the following general form:  
\begin{eqnarray}
k \frac {\partial U(\Phi^2,k)}{\partial k}&=& - \frac{k^6}{32 \pi^2}
   \left( \frac{4 N_c N_f}{k^2+M_q^2} - 3
   \frac{1}{k^2+M_\pi^2}-\frac{1}{k^2+M_\sigma^2} \right) \label{eq:Uk}\\
k \frac {\partial Z_\Phi(\sigma_0^2,k)}{\partial k}&=& - \frac{k^6}{16 \pi^2}
   Z_\Phi \left( \frac{4 N_c N_f}{(k^2+M_q^2)^3}\frac{G^2}{Z_q^2 Z_\Phi} + 4 \Lambda^2
    \frac{F_\pi^2}{(k^2+M_\pi^2)^2 \ (k^2+M_\sigma^2)^2}
    \right)\Big|_{\sigma_0^2}\label{eq:Zphik}\\
k \frac {\partial G(\sigma_0^2,k)}{\partial k}&=& - \frac{k^6}{16 \pi^2} 
    \frac{G^3}{Z_q^2 Z_\Phi}\left(\frac{6k^2+3M_q^2+3M_\pi^2}{(k^2+M_q^2)^2\
    (k^2+M_\pi^2)^2} -\frac{2k^2+M_q^2+M_\sigma^2}{(k^2+M_q^2)^2\ (k^2+M_\sigma^2)^2}\right)\Big|_{\sigma_0^2}\label{eq:Gk}\\
k \frac {\partial Z_q(\sigma_0^2,k)}{\partial k}&=& - \frac{k^6}{32 \pi^2}
   \frac{G^2}{Z_q Z_\Phi}
    \left(9 \frac{M_q^2+M_\pi^2}{(k^2+M_q^2)^2\
   (k^2+M_\pi^2)^2} -6 k^2
    \frac{(M_q^2-M_\pi^2)^2}{(k^2+M_q^2)^3\ (k^2+M_\pi^2)^3}
    \nonumber \right.\\
& &\left.  +  3 \frac{M_q^2+M_\sigma^2}{(k^2+M_q^2)^2\ (k^2+M_\sigma^2)^2}
    - 2k^2 \frac{(M_q^2-M_\sigma^2)^2}{(k^2+M_q^2)^3\
   (k^2+M_\sigma^2)^3}\right)\Big|_{\sigma_0^2}. \label{eq:Zqk}
\end{eqnarray}
The flow equations describe the evolution of the effective potential
$U$, the Yukawa coupling $G$, and the wave function renormalizations
$Z_\Phi$ and $Z_q$ under a change of the renormalization scale
$k$. 
The effective masses in the flow equations are given by
\begin{align}
M_q^2(\Phi^2,k)&= \frac{G^2}{Z_q^2}\Phi^2,\label{eq:MQ}\\
M_\pi^2(\Phi^2,k)&= \frac{2}{Z_\Phi}\frac{\partial U}{\partial
  \Phi^2}, \label{eq:MPI}\\
M_\sigma^2(\Phi^2,k)&= \frac{2}{Z_\Phi} \left( \frac{\partial U}{\partial
  \Phi^2}+2 \frac{\partial^2 U}{(\partial
  \Phi^2)^2} \Phi^2 \right),\label{eq:MSIGMA}\\
\end{align}
the effective pion decay constant $F_\pi(\Phi^2)$ by
\begin{align}
F_\pi^2&= Z_\Phi \Phi^2,\label{eq:Fpi(Phi)}
\end{align}
and the effective four-boson coupling $\Lambda(\Phi^2)$ by
\begin{align}
\Lambda&=\frac{2}{Z_\Phi^2} \frac{\partial^2 U(\Phi^2)}{(\partial
  \Phi^2)^2}=\frac{M_\sigma^2-M_\pi^2}{2\ Z_\Phi  \Phi^2}.\label{eq:Lambda(Phi)}
\end{align}
They depend on the field $\Phi^2$ and the scale $k$.
Evaluated at the vacuum expectation value $\Phi=\sigma_0$, the masses reduce to
the physical masses at the scale $k$
\begin{equation}
  m_q(k)=\left. M_q(\Phi^2,k) \right|_{\Phi^2=\sigma_0^2},  \quad
  m_\pi(k)=\left. M_\pi(\Phi^2,k) \right|_{\Phi^2=\sigma_0^2}, \quad
  m_\sigma(k)=\left. M_\sigma(\Phi^2,k) \right|_{\Phi^2=\sigma_0^2}.\label{physmasses}
\end{equation}
A flow equation for the vacuum expectation value $\langle\sigma\rangle=\sigma_0$ can be derived from
the minimum condition of the effective potential, 
\be
\frac{\partial}{\partial \sigma} U(\Phi^2, 
  \sigma, k)\bigg|_{\sigma=\sigma_0, \vec{\pi}^2=0} = 0.
\ee
Details of the derivation can be found in appendix \ref{app:flowsigma0}, here we
only quote the result
\be
k \frac{\partial}{\partial k} \sigma_0 &=& -\frac{2\ \sigma_0}{Z_{\Phi}
  M_\sigma^2(\sigma_0^2)} k\frac{\partial }{\partial k}
  U_0^\prime(\Phi^2,k) \bigg|_{\Phi^2=\sigma_0^2},\label{eq:min}
\ee
where $U_0(\Phi^2, k)$ denotes the symmetric part of the effective potential,
without the symmetry breaking term proportional to $\delta$, and the
prime denotes a derivative in the symmetric variable $\Phi^2$. Note
that the set of RG flow equations (\eqref{eq:Uk}-\eqref{eq:Zqk})
depends only on the symmetric variable $\Phi^2$. 
Explicit symmetry breaking introduced by the linear term $\delta \sigma$ affects only the initial 
condition $\sigma_0(k_{\scriptstyle{UV}})$ in eq.~\eqref{eq:min}.

\begin{figure}
\psfrag{GG}[]{${\rm [MeV]}$} 
\psfrag{k}[]{$k \; {\rm [MeV]}$} 
\psfrag{SS}[]{$\ \ \ \big| $} 
\psfrag{800}[]{$\ \ \ \ \ \ \ k_{\chi SB}$} 
\includegraphics[scale=1.3]{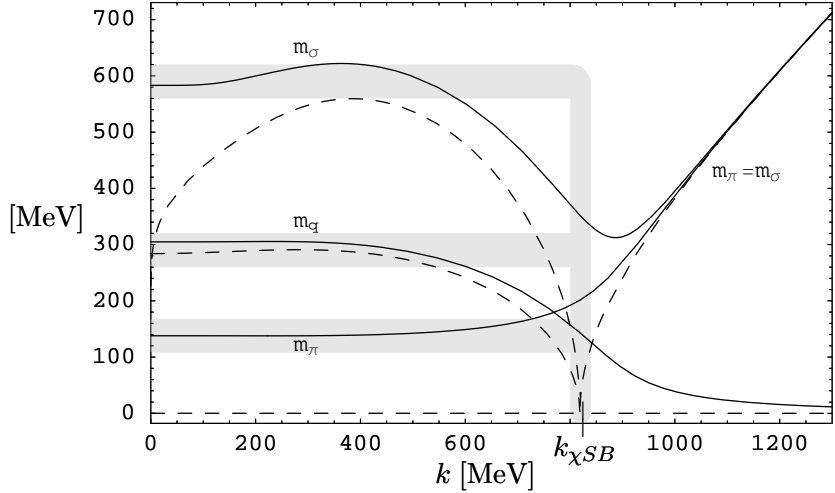}
\caption{The scale-dependent constituent quark mass $m_q$ and the  
meson masses
$(m_\sigma, m_\pi)$ as functions of the infrared cutoff parameter
$k$. Shown are results for simplified solutions of the RG flow
equations with constant wave function normalizations $Z_\Phi = Z_q
\equiv 1$ and constant Yukawa coupling $G$. Dashed lines show the
chiral limit $m_c \to 0$, while solid lines give the solutions with
explicit symmetry breaking. The grey bars indicate the bands in which
the masses are approximated as constant in the analytical calculation
of section IV.}
\label{fig:kflow}
\end{figure}
Two generic solutions of the renormalization group flow equations are
shown in Fig.~\ref{fig:kflow}. 
For zero current quark mass $m_c=0$, spontaneous chiral symmetry breaking sets in at the chiral symmetry breaking scale $k_{\chi SB}$ and then develops fully in the infrared.   
This picture is characterized by a non-vanishing quark mass $m_q$ and a 
sigma mass $m_\sigma\approx2m_q$ in the infrared. Below the chiral symmetry breaking scale, 
the evolution of the vacuum expectation value $\sigma_0$ quickly freezes.
For a finite current quark mass $m_c$, see full lines in Fig. 1. The
transition from the UV-dynamics to the IR-dynamics is smooth, the
meson masses never become zero.

\section{Analytical low energy expansions from the Renormalization Group}
\label{sec:analytical}

In this section, we show that the RG flow equations of the
bosonized NJL model generate the proper leading order pion contributions to physical observables. 
Due to the long-range fluctuations of the pions, a logarithmic term
(chiral log) appears in the quark mass expansion. The RG flow equations
generate such a term. Their full solution is best suited to sum up all
 long-range fluctuations, as we know from the physics of second order
 phase transitions. Since the RG flow equations go beyond an expansion in powers
of the coupling constants, they cannot be easily mapped onto a perturbative
expansion. However, we can identify terms with the same analytical behavior in
the lowest order pion mass, the parameter $M$, and make a direct comparison
to $\chi$PT results. In particular, the one-loop perturbative result can be
recovered from RG flow equations by replacing all running masses
by constant masses in lowest order \cite{Litim:2001ky}. 
Treating the small pion mass as constant in the RG flow equations is
justified only below the chiral symmetry breaking scale $k_{\chi SB}$,
where the pion exists as a low mass and low momentum degree of freedom and pion
fluctuations are the dominating effects. 
The chiral symmetry breaking scale $k_{\chi SB}$ in the RG approach 
appears as a UV cutoff. Although $k_{\chi SB}$ sets the scale for the logarithmic dependence on the pion mass parameter $M^2$, additional contributions may modify the value.

As a consequence of fixing the masses, the flow equations
\eqref{eq:Uk} and \eqref{eq:Zphik} decouple from equations
\eqref{eq:Gk} and \eqref{eq:Zqk}. This can be seen by substituting equations \eqref{eq:Fpi(Phi)} and \eqref{eq:Lambda(Phi)} in
equation \eqref{eq:Zphik} and replacing the ratio 
$\frac{G^2}{Z_q^2}$ by $\frac{m_q^2}{\sigma_0^2(k)}$. Therefore the evolution of $G$ and $Z_q$ is
not relevant for the solution of equations \eqref{eq:Uk} and
\eqref{eq:Zphik}. The system reduces to only two relevant
equations for the quantities we consider, one for
the effective potential $U$ and one for the wave function
renormalization $Z_\phi$,
\begin{eqnarray}
k \frac {\partial U}{\partial k}&=& - \frac{k^6}{32 \pi^2}
   \left( \frac{4 N_c N_f}{k^2+m_q^2} - 3
   \frac{1}{k^2+m_\pi^2}-\frac{1}{k^2+m_\sigma^2} \right) 
\label{eq:Uk2}\\
k \frac {\partial Z_\Phi}{\partial k}&=& - \frac{k^6}{16 \pi^2 \sigma^2_0}
    \left( \frac{4 N_c N_f}{(k^2+m_q^2)^3} m_q^2 +
    \frac{(m_\sigma^2-m_\pi^2)^2}{(k^2+m_\pi^2)^2 \ (k^2+m_\sigma^2)^2}
    \right).
\label{eq:Zphik2}
\end{eqnarray}
As pointed out in the introduction, the inclusion of the wave function
renormalization $Z_\phi$ in the RG flow is crucial to obtain the correct
behavior for the low $k$-dynamics.
We emphasize that the masses that enter on the right hand side of the flow
equations \eqref{eq:Uk2} and \eqref{eq:Zphik2} are set constant in the
$k$-flow and depend only on the parameter $M^2$. This is indicated by
using small letters for the masses.

To compare this approach with chiral perturbation theory, we expand the 
pion decay constant, the chiral quark condensate and the pion mass in 
the current quark mass $m_c$.
We distinguish between the chiral symmetric and the 
explicitly broken system by keeping the current quark mass $m_c$ in 
all expressions. The 
parameter $M^2$ in the $\chi$PT-expansion is directly related to $m_c$ by the 
Gell-Mann--Oakes--Renner relation. In lowest order we identify
\be
M^2 &=& 2 m_c B, \label{gmor2}
\ee
where $B$ is one of the low energy constants that appear in the chiral
Lagrangian \eqref{Lchpt}.

For explicit symmetry breaking 
the effective potential $U(\Phi^2, \sigma)$ depends on $M^2$ via the 
sigma field 
\begin{align} 
U(\Phi^2,\sigma) &= U_0(\Phi^2)-\delta \sigma, \label{Upar} \\
U_0(\Phi^2) &=
\frac{\lambda}{4}\left(\Phi^2-\Phi^2_0\right)^2,\label{U0par}
\end{align} 
where we expanded the potential up to fourth order. We parametrize the mesonic fields and the symmetry breaking parameter
as 
\begin{align}
\Phi=\left( \begin{array}{c} 
               \sigma\\ \vec \pi 
               \end{array} \right)\; {\rm and}\; \;
             \delta=\frac{m_{UV}^2}{g_{UV}} m_c.
\end{align}
The chiral and physical minima of the effective potential are defined by 
\begin{align} 
\Phi_0= \left( \begin{array}{c} 
               \phi_0\\ \vec 0 
               \end{array} \right)\ \ \  \rm{and}\ \ \   
\Phi_{phys}= \left( \begin{array}{c} 
               \sigma_0\\ \vec 0 
               \end{array} \right),
\end{align} 
which determines the pion decay constants in the chiral limit and the
physical case as  
\begin{align} 
f_{\pi0}=\sqrt{Z_\Phi} \ \phi_0 \quad {\rm and} \quad 
f_{\pi}=\sqrt{Z_\Phi} \ \sigma_0 .
\end{align} 
We will denote the difference between the physical pion 
decay constant and the decay constant in the chiral limit by $\Delta f_\pi\equiv f_\pi -f_{\pi0}$. The 
renormalized coupling constant is $\lambda_R=\frac{\lambda}{Z_\Phi^2}$.

Starting from the mass equations \eqref{eq:MQ}, \eqref{eq:MPI} and
\eqref{eq:MSIGMA}, we identify the $M^2$-dependence of the quark and
meson masses. 
For the pion mass we find
\begin{eqnarray} 
m^2_{\pi0}=&\frac{1}{Z_\Phi} \frac{\partial^2 U_0(\Phi^2)}{\partial \pi^2} \Big|_{\Phi_0}&=0, \\ 
m^2_{\pi}=&\frac{1}{Z_\Phi} \frac{\partial^2 U_0(\Phi^2)}{\partial 
  \pi^2} \Big|_{\Phi_{\mathrm{phys}}} 
&=2  \lambda_R f_{\pi0} \Delta 
f_\pi+{\cal O}(\Delta f_\pi^2). \label{mpi2} 
\end{eqnarray} 
As in chiral perturbation theory, the parameter $M$ is the
pion mass in lowest order of the chiral expansion 
\begin{eqnarray} 
m^2_{\pi}=M^2 +{\cal O}(M^4),
\end{eqnarray} 
which enables us to identify
\begin{align} 
M^2=2 \lambda_R \ f_{\pi0}\  \Delta f_\pi. \label{M} 
\end{align} 
Analogously, we derive
the expansions for the sigma mass and the constituent quark mass: 
\begin{align} 
m^2_{\sigma0}=&\frac{1}{Z_\Phi}\frac{\partial^2 U_0(\Phi^2)}{\partial\sigma^2} \Big|_{\Phi_0} 
=2 \lambda_R f_{\pi0}^2, \label{msigma0}\\ 
m^2_{\sigma} =&\frac{1}{Z_\Phi}\frac{\partial^2 U_0(\Phi^2)}{\partial\sigma^2} \Big|_{\Phi_{\mathrm{phys}}} 
             =m^2_{\sigma0}+3 M^2, \label{msigma}\\ 
m_{q0}=&\frac{G}{Z_q}|\Phi|\Big|_{\Phi_0}= \frac{G}{Z_q}\ 
\phi_0, \label{mq0}\\ 
m_{q}=&\frac{G}{Z_q}|\Phi|\Big|_{\Phi_{\mathrm{phys}}}= 
\frac{G}{Z_q \sqrt{Z_\Phi}} (f_{\pi0}+\Delta f_\pi) 
     =m_{q0}+A\ M^2, \label{mq} 
\end{align} 
 where the coefficient of the $M^2$-term of the constituent 
quark mass is given by 
 \begin{align} 
A=\frac{m_{q0}}{2 f_{\pi 0}^2\lambda_R}=\frac{m_{q0}}{m_{\sigma 0}^2}.\label{A}  
\end{align} 
Once the values $m_{\sigma 0}$
and $m_{q 0}$ for the sigma mass and the
constituent quark mass in the chiral limit are fixed, the
$M^2$-correction terms in $m_\sigma$ and $m_q$ enter 
the flow equations. Although these terms do not contribute to
the logarithmic corrections in $m_\pi^2$ and $f_\pi$, they influence
the $M^2$-corrections.

We first discuss the results for the pion decay constant $f_\pi=\sigma_0 \sqrt{Z_\Phi}$. To derive its flow we need the evolution equations for 
$\sigma_0$ and $Z_\Phi$. With the parametrization of 
the effective potential $U(\Phi^2,\sigma)$ (eq.~\eqref{Upar}),
we can express the evolution of the expectation value through the 
evolution of the effective potential. The expectation value satisfies 
\be 
 \frac{\partial}{\partial \sigma} U(\Phi^2, 
  \sigma)\bigg|_{\sigma=\sigma_0, \vec{\pi}^2=0}  =\frac{\partial 
    U_0(\Phi^2)}{\partial \Phi^2}\bigg|_{\Phi^2=\sigma_0^2}\, 2 \sigma_0 - \delta=U_0^\prime(\sigma_0^2)\, 2 \sigma_0 - \delta= 0,   
\label{eq:mincon} 
\ee 
since necessarily $\vec{\pi}=\vec{0}$. The prime denotes the derivative of 
$U_0(\Phi^2)$ with respect to $\Phi^2$.  
While the coefficient of the symmetry breaking term $\delta$ does not 
evolve under the RG flow, all other coefficients in the potential 
$U_0(\Phi^2)$ evolve, so that $\sigma_0=\sigma_0(k)$ is a function of the RG 
scale. Details of the derivation can be found in
appendix \ref{app:flowsigma0}. Combining eq.~\eqref{eq:mincon} with the flow equation for the
derivative coupling $Z_\Phi$, we find
\be 
k \frac{\partial}{\partial k} f_\pi^2 &=& \frac{k^6}{8 \pi^2} \left\{ 
\frac{1}{(k^2 + m_\pi^2)^2} +  
\frac{1}{(k^2 + m_\sigma^2)^2}  + \frac{1}{(k^2 + m_\pi^2) (k^2 + 
  m_\sigma^2)}\right. \nn\\ 
&&\;\;\;\;\;\; -\frac{3}{2} \frac{m_\pi^2}{m_\sigma^2} \left(\frac{1}{(k^2 + 
  m_\pi^2)^2} + \frac{1}{(k^2 + m_\sigma^2)^2} \right) \nn\\ 
&&\;\;\;\;\;\; \left .  -\frac{1}{2} 4 N_c N_f m_q^2 
\left(\frac{2}{m_\sigma^2} \frac{1}{(k^2+m_q^2)^2} + \frac{1}{(k^2 
  +m_q^2)^3} \right) \right \}\label{fpik} 
\ee 
Integrating this equation, we find a logarithmic correction to $f_\pi$ 
\begin{align} 
f_\pi &=f_{\pi0}\left( 1-\frac{M^2}{16 \pi^2 f_{\pi0}^2} \log 
\frac{M^2}{k_{\chi SB}^2}+{\cal  O}(M^2) \right).\label{f-pientw} 
\end{align} 
The structure of the result is exactly the
same as in $\chi$PT (cf. eq.~\eqref{Fpi}). Crucial for this result is the inclusion of the bosonic wavefunction renormalization, which enters at order $1/N_c$ and is not recovered in the standard large $N_c$-limit. While the generic form of the logarithmic term arises entirely from IR pion fluctuations and is universal, the scale inside the logarithm as well as all other analytic corrections depend on the detailed dynamics at larger scales and cannot be trusted in this simplified analytic approximation. 
These contributions have to be studied in a detailed numerical analysis of the non-perturbative flow equations performed in the next section.

Analogously to the pion decay constant, we can derive a flow equation
for the square of the pion mass $m_\pi^2$. As can be seen from
equations \eqref{eq:MPI} and \eqref{physmasses}, the physical pion mass is derived from the potential $U(\Phi^2)$,
\begin{align} 
m^2_\pi= \frac{2}{Z_\Phi}\frac{\partial U}{\partial
  \Phi^2}\Big|_{\Phi=\Phi_{\mathrm{phys}}}.
\end{align}
Differentiating the flow equation for the effective potential (\ref{eq:Uk2})
with respect to $\Phi^2$ and combining it with the equation for $Z_\Phi$ (eq. (\ref{eq:Zphik2}))
leads us to the flow equation for $m_\pi^2$
\begin{align} 
k \frac{\partial}{\partial k} m_\pi^2 &=\frac{2}{Z_\Phi}k
\frac{\partial U'(\sigma_0^2)}{\partial k}-\frac{2 \ U'(\sigma_0^2)}{Z_\Phi^2} k
\frac{\partial Z_\Phi}{\partial k}+\frac{2 U''(\sigma_0^2)}{Z_\Phi} k \frac{\partial \sigma_0^2}{\partial k}\nn\\
&=\frac{k^6 m_\pi^2}{16 \pi^2 f_{\pi}^2}\left(\left( -\frac{1}{2}+ \frac{3\ m_\pi^2}{2\ m_\sigma^2}\right)\left(\frac{1}{(k^2+m_\pi^2)^2}+\frac{1}{(k^2+m_\sigma^2)^2}\right)-\frac{2}{(k^2+m_\pi^2)(k^2+m_\sigma^2)}\right.\nn\\
&\left. \qquad \qquad + \frac{4 N_c N_f}{(k^2+m_q^2)^2}m_q^2\left(\frac{1}{k^2+m_q^2}+\frac{1}{m_\sigma^2}\right)\right).\label{mpik}
\end{align}
We remind the reader that in this equation all values on the right
hand side are considered constant in $k$ and in lowest order in the
chiral expansion, i.e. in particular $m_\pi^2\equiv M^2$ and
$f_\pi\equiv f_{\pi0}$. Similarly to the case for $f_\pi$, integration gives us the logarithmic correction to
$m_\pi^2$
\begin{align}
m_\pi^2 &=M^2\left( 1+\frac{M^2}{32 \pi^2 f_{\pi0} ^2} \log 
  \frac{M^2}{k_{\chi SB}^2}+ {\cal 
  O}(M^2)\right).\label{m-pientw}
\end{align}
As before for $f_\pi$, we also find for $m_\pi^2$ the correct logarithmic
behavior compared with $\chi$PT. Again we do not calculate the
contributions linear in $M^2$, which we will compute numerically in the following
section. To conclude, we emphasize that the RG approach to the NJL model shows
analytically the same low energy behavior as $\chi$PT for the two
observables $f_\pi$ and $m_\pi^2$ considered here. It should be mentioned
that the values of the zeroth order terms $f_{\pi 0}$ and $M^2$ depend
on the specific values of the integration boundary $k_{\chi SB}$. With
the approximation of constant masses we cannot obtain reasonable
zeroth order values with the same scale $k_{\chi SB}$ for $f_\pi$ and $m_\pi^2$. 
For completeness, we have also derived the chiral expansion of the
quark condensate $\left\langle \bar{q} q \right\rangle$ in appendix
\ref{app:qq}. It also reproduces the logarithmic term known from
$\chi$PT.

\section{Numerical evaluation of the Low Energy Constants}
\label{sec:numerical}

In this section we present numerical results for the pion
decay constant and the pion mass from the NJL model as functions of the current
quark mass which parametrizes explicit chiral symmetry
breaking. As demonstrated in the
previous section, we reproduce the logarithmic dependence
of $f_\pi$ and $m_\pi^2$ on the symmetry breaking
parameter, since the fluctuations of the pions are correctly
included. In chiral perturbation theory, the low energy coupling constants
of the next-to-leading order chiral Lagrangian manifest themselves as scales
$\Lambda_i$ in the logarithmic terms of the chiral expansion. These scales contain modifications due to
additional terms proportional to $\sim M^2$.

For the numerical evaluation, we expand each coupling of the effective
action (eq. \eqref{GAMMA}) in a
Taylor series around the physical minimum $\Phi_{\mathrm{phys}}=(\sigma_0(k),
\vec{0})$. As discussed in section \ref{sec:RG}, we take for the
wavefunction renormalization functions $Z_\Phi$ and $Z_q$ and the
Yukawa coupling $G$ only the first, $\Phi$-independent term into
account. Since the effective potential $U$
determines the vacuum expectation value $\sigma_0$ and therefore is
the most important coupling, we expand $U$ as follows: 
\be
U(\Phi^2, \sigma, k) &=& \sum_{i=1}^4 a_{i}(k)(\Phi^2 - \sigma_0(k)^2)^i - \delta \sigma
\label{eq:parapotential}
\ee 
where as usual $\Phi^2 = \sigma^2 +\vec{\pi}^2$.
The minimum condition for $\sigma_0(k)$,
\be
\frac{\partial}{\partial \sigma} U(\Phi^2, 
  \sigma, k)\bigg|_{\sigma=\sigma_0(k), \vec{\pi}^2=0} = 0,
\ee
gives as an additional constraint
\be
a_1(k) &=& \frac{\delta }{2 \sigma_0(k)}.
\label{eq:constraint}
\ee
We solve the coupled set of four flow equations
\eqref{eq:Uk}-\eqref{eq:Zqk} numerically for different values of
the UV cutoff and for a wide range of current quark masses $m_c$.
To obtain its bosonized form we perform a Hubbard-Stratonovich
transformation of the initial NJL model. In this bosonized Lagrangian the
kinetic term for the mesonic fields and higher meson interaction terms
are zero, so that the meson potential is characterized only by 
$m_{\scriptstyle{UV}}$. This gives the full set of initial
conditions at the UV scale:
\be
U_{\scriptstyle{UV}}(\Phi^2,\sigma)&=& \frac{1}{2}
m^2_{\scriptstyle{UV}}\Phi^2-
\frac{m_{\scriptstyle{UV}}^2}{G_{\scriptstyle{UV}}}m_c \; \sigma\nn\\
Z_{\Phi, \scriptstyle{UV}} &\equiv& 10^{-9}\nn\\
G_{\scriptstyle{UV}}&\equiv&1\\
Z_{q, \scriptstyle{UV}} &\equiv&1\nn
\ee
For a given cutoff $\Lambda_{\scriptstyle{UV}}$, we determine the parameter
$m_{\scriptstyle{UV}}$ and the physical current quark mass $m_{\mathrm{phys}}$ by requiring that the values $f_\pi = 92.4 \pm 0.3 \;\mathrm{MeV}$ and $m_\pi = 138.0\pm 1.0
\; \mathrm{MeV}$ are reproduced within these tolerances. 
This condition is the most natural choice and
allows a calculation independent from $\chi$PT.
As a consequence, the value of the physical current quark mass
$m_{\mathrm{phys}}$ varies with the UV cutoff $\Lambda_{\scriptstyle{UV}}$. 
To compare our results with chiral perturbation theory it is
necessary to connect our symmetry breaking
parameter $m_c$ with the parameter $M^2$ used in $\chi$PT. As already
explained in section \ref{sec:analytical}, we set
\be
M^2 &=&2 m_c B.\nonumber
\ee

\begin{figure}[t]
\psfrag{M}[]{$\frac{m_c}{m_{\mathrm{phys}}}$} 
\psfrag{F}[]{\hspace{.9cm} $f_{\pi}$[${\rm MeV}$] \hspace{.9cm} }
\center{\includegraphics[scale=1.45]{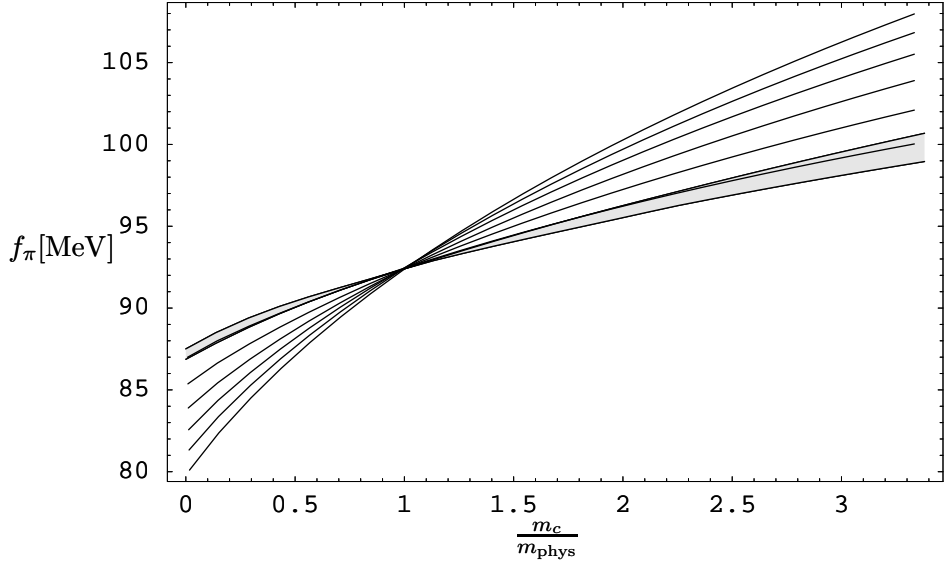}} 
\caption{\label{fig:2a} The pion decay constant $f_\pi$ is shown as a function of the
scaled current quark mass $m_c/m_{\mathrm{phys}}$ for different values
of the UV cutoff. The shaded region is given by $\chi$PT, obtained from
ref.~\cite{Leutwyler:2001}. The individual lines correspond to the full numerical RG results for
different cutoffs from $\Lambda_{UV} = 1.5 \; \mathrm{GeV}$
down to  $\Lambda_{UV} = 1.0 \; \mathrm{GeV}$ in steps of
$0.1 \; \mathrm{GeV}$. The curve with the largest slope belongs to
$\Lambda_{UV} = 1.5 \; \mathrm{GeV}$, the curve with the
smallest slope corresponds to $\Lambda_{UV} = 1.0 \;
\mathrm{GeV}$. All curves are fixed to agree at the physical point ($m_c/m_{\mathrm{phys}} = 1$), which corresponds to $f_\pi = 92.4 \;
\mathrm{MeV}$.}
\end{figure}

\begin{figure}[h]
\psfrag{m}[]{ $\frac{m_c}{m_{\mathrm{phys}}}$} 
\psfrag{M2}[]{$m_{\pi}^2$[${\rm MeV}^2$] \hspace{.5cm}} 
\center{\includegraphics[scale=1.45]{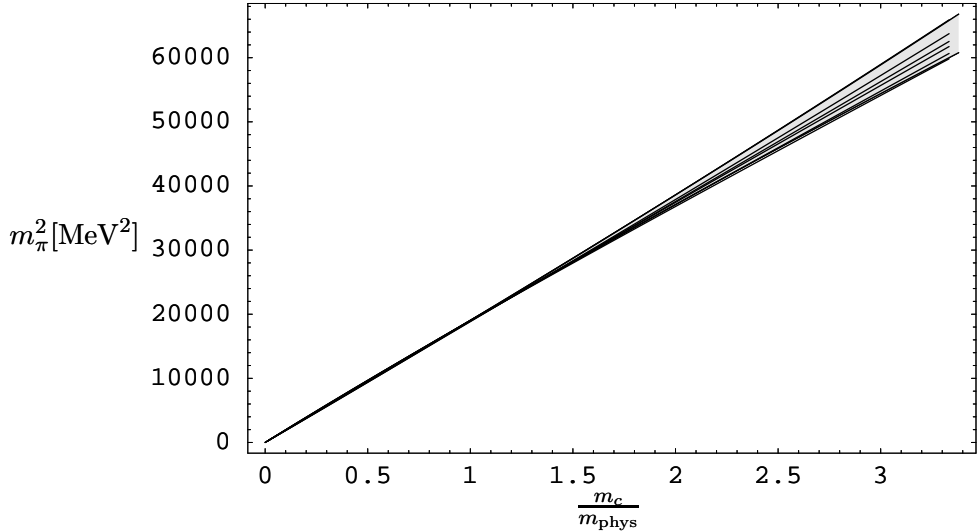}} 
\caption{\label{fig:2b} The pion mass is shown as function of the scaled
  current quark mass
  $m_c/m_{\mathrm{phys}}$ for different UV cutoffs $\Lambda_{UV}$. The shaded
  region is given by $\chi$PT, obtained from
  ref. \cite{Leutwyler:2001}. The lines correspond to
  the RG results for different cutoff choices, from $\Lambda_{UV}= 1.5\;
  \mathrm{GeV}$ (smallest slope) down to $1.0\; \mathrm{GeV}$ (largest slope)
  in steps of $0.1\; \mathrm{GeV}$.}
\end{figure}

In  Figs.~\ref{fig:2a} and \ref{fig:2b}, we compare our results
for $f_\pi$ and $m_\pi^2$ as functions of the current quark mass $m_c$
with those of $\chi$PT, obtained from \cite{Leutwyler:2001}.
We use a scaled current quark mass, since the absolute value of the quark mass does not have an immediate physical meaning and depends on a chosen scale.   
In ref.~\cite{Leutwyler:2001}, the $\chi$PT results are
plotted against the dimensionless variable $m_c/m_s$, where $m_c=m_u=m_d$ is
the mass of the u- or d-quark, and $m_s$ is the mass of the strange
quark. The physical pion decay constant and pion mass are obtained for
$m_c/m_s=1/26$. 

In order to be able to compare our results with $\chi$PT, we introduce a scale $m_{\mathrm{phys}}$ such that $m_c/m_{\mathrm{phys}}=1$ for the
physical value $f_\pi=92.4$ MeV, and we rescale the $\chi$PT results accordingly.
Over the considered range of UV cutoffs,
the value of $m_{\mathrm{phys}}$ varies from $m_{\mathrm{phys}} = 12.9 \; \mathrm{MeV}$ for $\Lambda_{\scriptstyle{UV}} = 1.0 \; \mathrm{GeV}$ to $m_{\mathrm{phys}} = 6.0\;
\mathrm{MeV}$ for $\Lambda_{\scriptstyle{UV}} = 1.5
\; \mathrm{GeV}$.
By construction, the RG gives the physical value for $f_\pi$ at
$m_c = m_{\mathrm{phys}}$ for all UV cutoffs
$\Lambda_{\scriptstyle{UV}}$. As shown in Fig.~\ref{fig:2a}, the curves
$f_\pi(\frac{m_c}{m_{\mathrm{phys}}})$ become steeper with increasing cutoff $\Lambda_{\scriptstyle{UV}}$.
For the cutoff value $\Lambda_{\scriptstyle{UV}}=
(1.00\pm 0.05)$ GeV, the RG result falls within the band given by $\chi$PT.
In the chiral limit $m_c \rightarrow 0$, the pion decay constant
from the RG and from $\chi$PT do not necessarily match. Depending on the value of the UV cutoff, our result for $f_{\pi 0}=f_\pi(m_c\to 0)$ varies between $79\;\mathrm{MeV}$ for a cutoff of $\Lambda_{\scriptstyle{UV}} = 1.5 \; \mathrm{GeV}$ and $87\; \mathrm{MeV}$ for a cutoff of  $\Lambda_{\scriptstyle{UV}}
= 1.0 \; \mathrm{GeV}$. In $\chi$PT, the chiral limit of the pion
decay constant is $f_{\pi 0} = 86.2 \pm 0.5 \;\mathrm{MeV}$~\cite{Colangelo:2003hf}.

We note that the slope of $f_\pi$ as a function of $m_c/m_{\mathrm{phys}}$ increases with
increasing value of $\Lambda_{\scriptstyle{UV}}$.
In the NJL model, this is due to
the increase of $m_{\scriptstyle{UV}}$ with the cutoff $\Lambda_{\scriptstyle{UV}}$, which can be seen
from Tab.~\ref{tab:fitconst}. As a result, for a given current quark mass
$m_c$, the linear symmetry breaking term $\delta =
\frac{m_{\scriptstyle{UV}}^2}{G_{\scriptstyle{UV}}} \; m_c$ also
increases with the cutoff. This leads to the increase of the slope
of $f_\pi(\frac{m_c}{m_{\mathrm{phys}}})$ with $\Lambda_{\scriptstyle{UV}}$ observed in
Figure~\ref{fig:2a}.

The results for the pion mass $m_\pi^2$ are shown in
Fig.~\ref{fig:2b} for the same set of UV cutoffs. In contrast to the pion decay constant, the slope
of $m_\pi^2$ as a function of the symmetry breaking parameter decreases with
increasing $\Lambda_{\scriptstyle{UV}}$.
The RG result falls within the band of the $\chi$PT result for a much larger
range of cutoffs $\Lambda_{\scriptstyle{UV}}$, compared with the result of the
pion decay constant. Since the
corrections to $m_\pi^2$ are already of order $M^4$, and thus one order
higher than the corrections to $f_\pi$, $m_\pi^2$ increases almost
linearly with $m_c/m_{\mathrm{phys}}$.
We find that for $\Lambda_{\scriptstyle{UV}} \approx 1.0
\; \mathrm{GeV}$ the results for $f_\pi$ and $m_\pi^2$ are consistent with those of $\chi$PT, see Figs.~\ref{fig:2a} and \ref{fig:2b}.
However, if one does not consider any additional physical observable, there
is {\it a priori} no reason to favor any particular value of
the UV cutoff for the NJL model, as long as the cutoff varies
between $1\  \rm{GeV} < \Lambda_{\scriptstyle{UV}} < 1.5\ \rm{GeV}$. A
cutoff much smaller than 1 GeV is not justified, since the
phenomenological $\Lambda_4$-parameter is around 1 GeV. Consistent
with this, ref.~\cite{Schwenzer:2005} finds that for a cutoff slightly below
$\Lambda_{\scriptstyle{UV}} < 1$ GeV the RG equations
cannot reproduce the physical $f_\pi$. A much larger cutoff would
extend the effective four-fermion interaction into a region where
dynamical gluon effects become important.

\begin{figure}[b]
\psfrag{1.35}[]{\hspace{.6cm} $\frac{f_{\pi}}{f_{\pi0}}$} 
\psfrag{x}[]{\shortstack[]{\vspace{-.9cm}$\frac{m_c}{m_{\mathrm{phys}}}$}}
\psfrag{M}[]{$\frac{m_{\pi}^2}{M^2}$ \hspace{.9cm}}
\center{\includegraphics[scale=1.0]{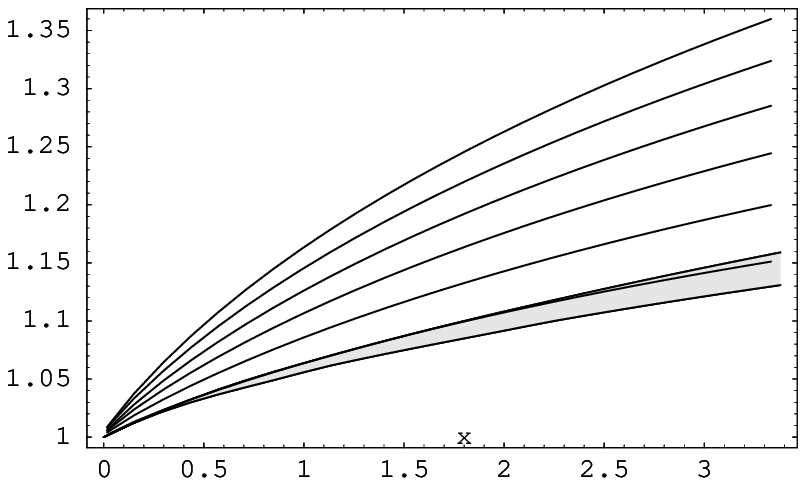} \includegraphics[scale=1.0]{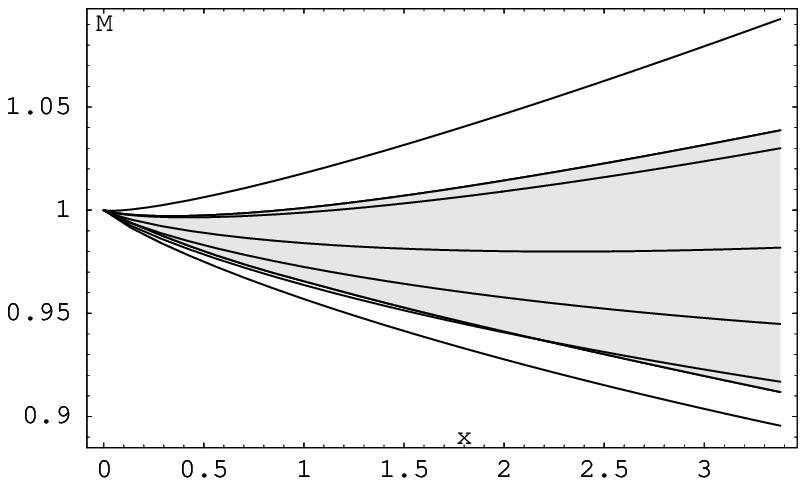}}
\caption{\label{fig:3a} We plot the ratios $f_\pi/f_{\pi 0}$ and $m_\pi^2/M^2$ as  
functions of
the scaled current quark mass $m_c/m_{\mathrm{phys}}$. The $\chi$PT
results of ref.~\cite{Leutwyler:2001} give the shaded regions, the RG flow
equations yield the individual lines. For $f_\pi / f_{\pi 0}$, the RG
results are shown from $\Lambda_{UV} = 1.5 \; \mathrm{GeV}$
(top line) to  $\Lambda_{UV} = 1.0 \; \mathrm{GeV}$ (bottom
line), and for $m_\pi^2/M^2$, from $\Lambda_{UV} = 1.0 \;  
\mathrm{GeV}$
(top line) to  $\Lambda_{UV} = 1.5 \; \mathrm{GeV}$ (bottom
line), in steps of $0.1 \; \mathrm{GeV}$.}
\end{figure}
While the leading-order behavior of $f_{\pi 0}$ and the slope of
$m_\pi^2$ as a function of $m_c$ depend
on the implementation of the symmetry breaking term in the model Lagrangian,
the ratios $\frac{f_\pi}{f_{\pi0}}$
and $\frac{m_\pi^2}{M^2}$ are more indicative of pion fluctuations
(cf. Figs.~\ref{fig:3a}). They are determined by the
pion mass in lowest order $M$ and the low energy constants $\Lambda_3$ and $\Lambda_4$.

The pion mass $m_\pi^2$ is normalized
with $M^2 = 2 B m_c$, where the value for $2 B$ has been obtained from the
leading order fit to the pion mass. In the chiral limit, we have
\be
\lim_{m_c \to 0} \frac{m_\pi^2}{M^2} = 1.
\ee
With this normalization, corrections to the linear $m_c$-dependence can be
compared with results of $\chi$PT~\cite{Leutwyler:2001}, see shaded
regions in Figs.~\ref{fig:3a}.
In general, the cutoff sensitivity is enhanced in
the expansion around $m_c=0$. At the maximum value of our calculation 
$m_c/m_{\mathrm{phys}}=3.4$ it amounts to 20 $\%$.

In order to obtain the low energy constants, we fit the pion decay constant $f_\pi(m_c)$ and the
pion mass $m_\pi^2(m_c)$ (given in Figs.~\ref{fig:3a}) as
functions of $m_c$ to the parametrization from $\chi$PT at one-loop order \cite{Leutwyler:2001}
\be
f_\pi &=& f_{\pi 0}\left( 1  - \frac{1}{16 \pi^2} \frac{2 B m_c }{f_{\pi 0}^2} \log \frac{2 B
  m_c}{\Lambda_4^2} + {\mathcal O}(m_c^2)\right)\label{fpifit}\\
\frac{m_\pi^2}{m_c} &=& 2 B\left(1 + \frac{1}{32 \pi^2} \frac{2 B
  m_c}{f_{\pi 0}^2} \log \frac{2 B m_c}{\Lambda_3^2} + {\mathcal O}(m_c^2) \right).\label{mpifit}
\ee
We use current quark masses up
to $m_c = 20 \; \mathrm{MeV}$, which corresponds to pion masses up to
$\sim 250  \; \mathrm{MeV}$, depending on the exact value of the UV
cutoff.

\begin{table}[t]
\begin{center}
\begin{tabular}{||ccc||cccc|cc||}
\hline
\hline
$$  & Input [MeV]& $\;\;$ & $
\;\;$$\;\;$ &$\;\;\;$Output [MeV]& &
$\;\;$& \;\;\;\;\;\;\;Output& \\
\hline
$\;\;\;\Lambda_{\scriptstyle{UV}}\;$  & $\;m_{\scriptstyle{UV}}\;$& $\;m_{\mathrm{phys}}\;\;$ & $
\;\;\;f_{\pi 0}$\;\;\; & $ M(m_{\mathrm{phys}})$ & $\;\;\Lambda_3\;\;$ & $\;\;\;\;\;\;\;\Lambda_4\;\;\;\;\;\;\;$&$\;\;\;\;\; \bar{l}_3\;\;\;\;\;\;\;$& $\bar{l}_4\;\;$\\
\hline
$1000$ & $233.2$ & $12.9$ & $86.9$ & $136.3$ & $
\phantom{2} 43.3 $ & $
1303.5 $ &${\scriptstyle{-}}2.29\;\;\;\;\;\;$ &  $ 4.52 \;\;$\\
$1100$ & $265.3$ & $10.6$ & $85.1$ & $137.2$ & $
\phantom{1} 147.2 $ & $
1297.6 $ &$\phantom{1} 0.14\;\;\;\;\;\;$ &  $ 4.49\;\;$\\
$1200$ & $296.0$ & $\phantom{1} 9.0$ & $83.5$ & $138.5$ & $
\phantom{1} 346.7$ & $ 1261.1 $ &$\phantom{1} 1.72\;\;\;\;\;\;$ &  $ 4.42\;\;$\\
$1300$ & $325.8$ & $\phantom{1} 7.8$ & $82.1$ & $139.8$ & $
\phantom{1} 624.5 $ & $ 1245.4 $ &$\phantom{1} 2.99\;\;\;\;\;\;$ & $ 4.37\;\;$\\
$1400$ & $355.0$ & $\phantom{1} 6.8$ & $80.7$ & $140.4$ & $
\phantom{1} 931.0 $ & $ 1247.8 $&$\phantom{1} 3.78\;\;\;\;\;\;$ & $ 4.37 \;\;$\\  
$1500$ & $383.8$ & $\phantom{1} 6.0$ & $79.4$ & $141.0$ & $1217.0 $ &
$1261.3 $ &$\phantom{1} 4.31\;\;\;\;\;\;$ & $ 4.38\;\;$\\
\hline
\hline
\end{tabular}
\end{center}
\caption{\label{tab:fitconst} Values for the pion decay constant in
  the chiral limit $f_{\pi 0}$, $M(m_{\mathrm{phys}})$, the scale-independent low
  energy constants $\Lambda_3$ and $\Lambda_4$ and the scale-dependent
  quantities $\bar{l}_3$ and $\bar{l}_4$ at scale
  $M(m_{\mathrm{phys}})$ from the fits to the RG
  results, dependent on the UV values $\Lambda_{\scriptstyle{UV}}$ and
  $m_{\scriptstyle{UV}}$. The values of $m_{\mathrm{phys}}$ are determined by the
  condition that our result should reproduce the physical values of
  $f_\pi$ and $m_\pi$ at $m_{\mathrm{phys}}$.}
\end{table}
In Table~\ref{tab:fitconst}, we show the pion decay
constant in the chiral limit $f_{\pi 0}$ and the leading term of the
pion mass $M(m_{\mathrm{phys}})\equiv \sqrt{2 B m_{\mathrm{phys}}}$ as
functions of the UV cutoff $\Lambda_{\scriptstyle{UV}}$,
$m_{\scriptstyle{UV}}$ and $m_{\mathrm{phys}}$. As explained before, a fixed
cutoff $\Lambda_{\scriptstyle{UV}}$ leaves the two free parameters
$m_{\scriptstyle{UV}}$ and $m_{\mathrm{phys}}$, which are then fixed from $m_\pi^2$ and $f_\pi$.
The different values from Tab.~\ref{tab:fitconst} for different
cutoffs lead to the following intervals for  $f_{\pi 0}$ and $M(m_{\mathrm{phys}})$. We find
over the range of UV cutoffs considered
\be
79.4 \ \mathrm{MeV}   & \le f_{\pi 0} \le & \phantom{1}86.9 \ \mathrm{MeV},\label{fpi0value}\\
136.3 \ \mathrm{MeV} &\le M(m_{\mathrm{phys}})  \le & 141.0 \ \mathrm{MeV}.
\ee  
Fits to our full numerical RG results with equations (\ref{fpifit}) and (\ref{mpifit}) give
the low energy constants $\Lambda_3$ and $\Lambda_4$, which are also
listed in  Tab.~\ref{tab:fitconst}. We extract from the table the ranges
\be
0.04 \; \mathrm{GeV} & \le \Lambda_3 \le   1.22 \; \mathrm{GeV}, & -2.29 \le  \bar l_3 \le 4.31,\\
1.25 \; \mathrm{GeV} & \le \Lambda_4 \le 1.30  \; \mathrm{GeV}, &  \phantom{-}4.37 \le \bar l_4 \le 4.52 \label{lambda4value}.
\ee
These ranges are obtained by varying the UV cutoff between the 
maximal and minimal values in the model, determined by 
the momentum region for which the
NJL model can be considered as a suitable phenomenological description. 
Changing the cutoff, one finds from Table~\ref{tab:fitconst} for example a large value for $f_{\pi 0}$ correlated with a small value of $\Lambda_3$.
The values of the low energy constants used in $\chi$PT
\cite{Leutwyler:2001} are
\be
0.08 \; \mathrm{GeV} & \le \Lambda_3^{\scriptstyle{\chi PT}} \le   1.99 \; \mathrm{GeV}, & \phantom{-}0.5 \le  \bar l_3^{\scriptstyle{\chi PT}} \le 5.3,\\
1.12 \; \mathrm{GeV} & \le \Lambda_4^{\scriptstyle{\chi PT}} \le 1.40  \; \mathrm{GeV}, &  \phantom{-}4.2 \le \bar l_4^{\scriptstyle{\chi PT}} \le 4.6.
\ee
The constant $\Lambda_4$ ($\bar l_4$) reflects the correction of the leading
order result of $f_\pi$ due to quantum fluctuations, and thus is
less dependent on the particular implementation of chiral symmetry
breaking than $\Lambda_3$ ($\bar l_3$), which determines the next-to-leading order
term of the expansion of $m_\pi^2$.

\begin{figure}[ht!]
\psfrag{x}[]{ $\frac{m_c}{m_{\mathrm{phys}}}$} 
\psfrag{M}[]{${\rm [MeV]}\; \; \;\;$}
\psfrag{MPI}[]{ $m_\pi$} 
\psfrag{MQ}[]{$m_q$}  
\center{\includegraphics[scale=1.3]{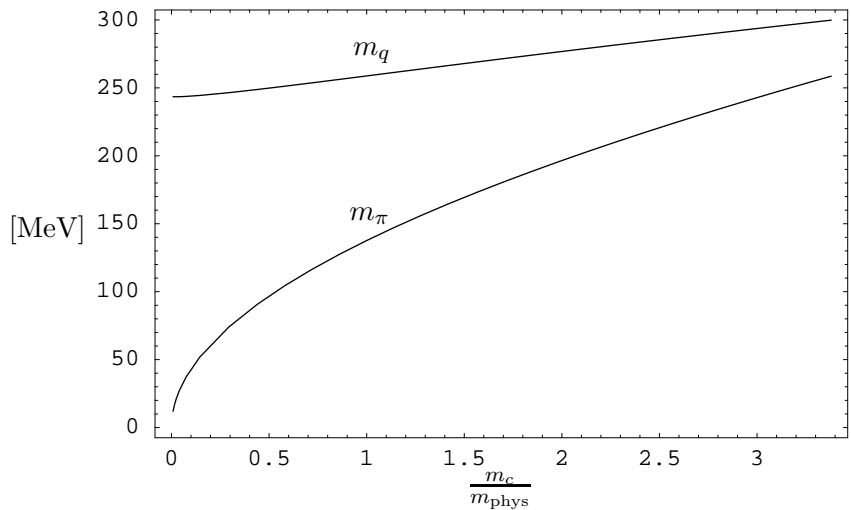}} 
\caption{\label{fig:4} Comparison between the pion mass $m_\pi$ and the
  constituent quark mass $m_q$, obtained from the RG flow equations, as functions of
  the scaled current quark mass $m_c/m_{\mathrm{phys}}$. A large mass gap exists between
  $m_\pi$ and $m_q$ for small current quark masses
  $m_c/m_{\mathrm{phys}}$. The plot is made exemplarily for a UV cutoff $\Lambda_{\scriptstyle{UV}}= 1.0\; \mathrm{GeV}$.}
\end{figure}
Deviations between the results from $\chi$PT and the RG treatment of the
bosonized NJL model are to some degree expected, in particular for large
values of the pion mass. In the NJL model the low momentum
regime contains free constituent quarks with masses of roughly $\sim
300 \; \mbox{MeV}$ (see Fig.~\ref{fig:4}) for realistic values of the pion mass and the pion
decay constant. In contrast to the assumption in $\chi$PT, the mass gap
between the light pseudo-Goldstone boson $m_\pi$ and the more massive
constituent quark $m_q$ becomes smaller for larger current quark masses, and
the dominance of the pion field fluctuations is lost. This suggests
that effects of quark loops or higher $(\bar{q} q)$ bound states
may be enhanced for larger $m_c/m_{\mathrm{phys}}$.

\begin{figure}[t]
\psfrag{x}[]{\shortstack[]{{\vspace{-.8cm} $\frac{m_c}{m_{\mathrm{phys}}}$}}}
\psfrag{M}[]{}
\psfrag{H}[]{$\scriptstyle{\Lambda_{UV}}=
  1.5\; \mathrm{GeV}$}
\psfrag{L}[]{$\scriptstyle{\Lambda_{UV}}=
  1.0\; \mathrm{GeV}$}
\psfrag{MMM}[]{$m_\pi$ \hspace{.9cm}} 
\psfrag{500}[]{\hspace{.5cm}$m_\pi$} 
\psfrag{130}[]{\hspace{.4cm}$f_\pi$} 
\includegraphics[scale=1.0]{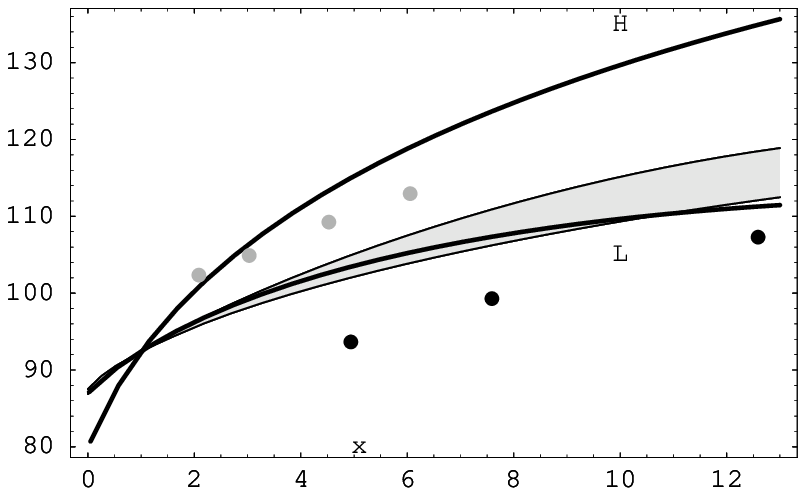} \includegraphics[scale=1.0]{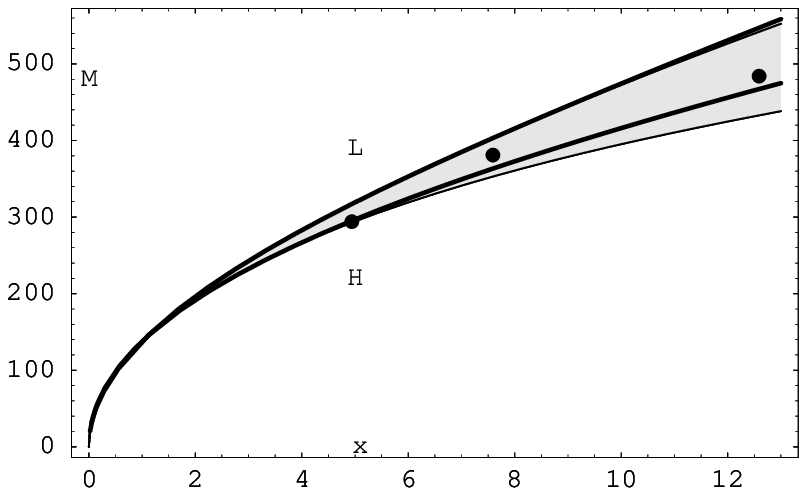}
\caption{\label{fig:5} The plots show the pion decay
  constant $f_\pi$ and the pion mass $m_\pi$ for 
  large values of the scaled quark masses $m_c/m_{\mathrm{phys}}$ for
  the two cutoff choices $\Lambda_{\scriptstyle{UV}}= 1.0\;
  \mathrm{GeV} (m_{\mathrm{phys}}=12.9\; \mathrm{MeV})$ and $\Lambda_{\scriptstyle{UV}}= 1.5\;
  \mathrm{GeV} (m_{\mathrm{phys}}=6.0\; \mathrm{MeV})$.
  The shaded regions represent the results of $\chi$PT, while the bold curves give the result from RG 
  flow equations. The curves are
  shown up to a quark mass $m_c\sim170\; \mathrm{MeV}$ and
  $m_c\sim80\; \mathrm{MeV}$, respectively.
  Results from lattice calculations with Wilson fermions from
  \cite{Luscher:2005mv} are plotted for comparison and represented by
  the bold dots, while the grey dots give results from
  \cite{Davies:2004} with staggered fermions.}
\end{figure}

Therefore we have investigated the range of
$m_c/m_{\rm phys}$ up to approximately 13. For cutoffs $\Lambda_{{\scriptstyle
    UV}}=1.0$ GeV and  $\Lambda_{{\scriptstyle UV}}=1.5$ GeV, this amounts to current quark masses up to $m_c \sim
170$ MeV and $m_c \sim
80$ MeV, respectively. The results for $f_\pi(m_c)$ and $m_\pi(m_c)$ are shown in
Figs.~\ref{fig:5}. For $\Lambda_{{\scriptstyle UV}}=1.0$ GeV, the expected variations for large quark masses do not
leave the band predicted by $\chi$PT \cite{Leutwyler:2001}. For $\Lambda_{{\scriptstyle UV}}=1.5$ GeV, the value of $f_\pi$ for large quark masses deviates more strongly from $\chi$PT. In the same figure, we show results of 
recent lattice results with Wilson fermions, taken from
\cite{Luscher:2005mv}, which were obtained with a  new pre-conditioned
Hybrid Monte-Carlo algorithm, and from \cite{Davies:2004} with
staggered fermions.

While the description of these lattice results does not require chiral
logarithms, they do not rule out such terms, either. A comparison of
large quark-mass lattice simulations with the curves of
Fig.~\ref{fig:5} allows an extrapolation of lattice data to small
quark masses, as long as they lie in the region covered by our
calculation and their slope is compatible with our results. A set of
values for $f_\pi$ and $m_\pi$ for large quark masses from our calculation can be found in table~\ref{tab:largemc} in appendix~\ref{app:table}.

\section{Conclusions}
\label{sec:conclusions}

We have studied the quark mass dependence  of low energy observables within the NJL model, using renormalization group flow equations. We have shown that these equations analytically generate the non-analytic terms in the chiral expansion known from chiral perturbation theory.
Our result confirms that the pion contributions to the low energy constants are treated correctly in the RG flow equations. For this it is essential to consider the renormalization of
the kinetic term of the pion fields. 
While it is difficult in general to map non-perturbative RG results on
a perturbative expansion, it is much easier to identify non-analytic
parts. To derive these results, we treat the pion mass as a constant
in the RG flow equation below the chiral symmetry breaking
scale. Integrating the RG flow is then equivalent to performing a
perturbative one-loop calculation.  The resulting chiral logarithms $
\sim M^2 \log M^2$ are accompanied by an analytic power series in
$M^2$, where $M$ is the pion mass to lowest order.

Going beyond the analytical results, we find that the numerical solution
of the full set of the RG flow equations gives low energy constants which are
compatible with the phenomenological values from $\chi$PT. 
As input to these equations, in the bosonized NJL model one needs the
triple of an UV cutoff $\Lambda_{\scriptstyle{UV}}$, a mass parameter
$m_{\scriptstyle{UV}}$ and a current quark mass $m_{\mathrm{phys}}$. 
For a given value of the UV cutoff, the remaining input parameters are adjusted to reproduce the physical pion
decay constant $f_\pi$ and the pion mass $m_\pi^2$ at the physical
point. Therefore there is some freedom in choosing such a triple. 
Output of the RG approach are the values of the pion decay
constant in the chiral limit $f_{\pi 0}$, the pion mass in lowest
order $M(m_{\mathrm{phys}}) = \sqrt{2 B m_{\mathrm{phys}}}$, 
and the two low energy constants $\Lambda_3$ and
$\Lambda_4$ (see Table~\ref{tab:fitconst}). When the UV cutoff
$\Lambda_{\scriptstyle{UV}}$ is varied in a reasonable range $1.0\  {\rm
  GeV} <\Lambda_{\scriptstyle{UV}} < 1.5\  {\rm
  GeV}$, the chiral limit of the pion decay constant varies as  $79.4 \ \mathrm{MeV} \le  f_{\pi 0} \le
86.9 \ \mathrm{MeV}$ and the pion mass in
lowest order as $136.3 \ \mathrm{MeV} \le  M(m_{\mathrm{phys}}) \le 141.0 \
\mathrm{MeV}$. The corresponding low energy constants are $0.04 \ \mathrm{GeV} \le \Lambda_3 \le 
1.22 \ \mathrm{GeV}$ and 
$1.25 \ \mathrm{MeV} \le \Lambda_4 \le 1.30 \ \mathrm{GeV}$. For the renormalization
scale-independent quantities we find $- 2.29 \le \bar l_3
\le 4.31$ and $4.37 \le \bar l_4 \le 4.52$ at the mass scale
$M(m_{\mathrm{phys}})$, see equation \eqref{liconstants}.

The NJL model lacks confinement, and therefore our model contains free quarks. As a
consequence, the mass gap between the
light pseudo-Goldstone bosons and the other hadronic states is reduced
to a mass gap between the constituent quarks and pions.
Nevertheless, our results show that the values of the low energy constants
which we obtained from the model are compatible with chiral
perturbation theory. On the other hand it is well known \cite{Gasser:1983yg} that the linear $\sigma$-model without quarks cannot reproduce chiral perturbation theory: Quark
loops are essential to mimic effects of higher mass meson resonances
like the $\rho$ meson.

The topic of low energy constants in phenomenological models has been
addressed in many publications, see e.g. \cite{Schuren:1991sc,
  RuizArriola:1991gc, RuizArriola:1991bq, Klevansky:1992qe,
  Schuren:1993aj, Mueller:1994dh, Hippe:1995hu, Bijnens:1995ww,
  Jungnickel:1997yu, Llanes-Estrada:2003ha}.
Our results complement an earlier study \cite{Jungnickel:1997yu} that
aimed at the low energy expansion within the linear $\sigma$-model and
gave estimates for the low energy constants $L_4$ - $L_8$ in the
$SU(3) \otimes SU(3)$ parametrization, which correspond to $l_1$ -
$l_3$, $l_5$ and $l_6$ in the $SU(2) \otimes SU(2)$ case. These
results showed a qualitative agreement with phenomenology, but relied
on a mean field approximation and introduced an arbitrary matching
scale. Since these assumptions are not necessary in our approach, it
would be an interesting  challenge to study also the low energy
expansion of $\chi$PT at NLO. This can in principle be done without
additional assumptions by an extension of our effective action to NNLO
in the derivative expansion.

The results presented here demonstrate that the NJL model gives a
reasonable description of the infrared chiral dynamics and its quark
mass dependence. By construction, it is based on the same
representation of the chiral flavor symmetry as QCD. Therefore, it involves degrees of freedom that go beyond the low energy dynamics of the pions that are fixed by the symmetry. 
The requirement that a model must be consistent with the physical low
energy constants provides constraints on the full set of possible dynamical theories at large momentum scales.
We find that including quark effects, the results from the model are compatible with $\chi$PT, whereas a linear $O(N)$-model without quarks is not \cite{Gasser:1983yg}.
In this respect the NJL model passes an important test for
applicability to chiral physics in the
non-perturbative regime. This is also relevant for dense quark matter in
the deconfined phase. Due to the difficulty of performing lattice gauge
simulations in this regime, the NJL model is widely used as a tool for investigations of
color-superconducting quark matter, see e.g. \cite{Buballa:2003qv}, and its phase structure, which is sensitive to the quark masses \cite{Ruster:2005jc}.

Finally, a quark mass expansion is essential to extrapolate lattice gauge simulations to the physical regime. The lowest-order expansion of chiral perturbation theory will be sufficient as long as the lattice masses are sufficiently small, but it may become unreliable for large masses which are required for studies with chiral fermions \cite{Kaplan:1992bt, Neuberger:1997fp}. 
A description that explicitly includes quark dynamics beyond the infrared regime may give a better account of the large quark-mass behavior. 
We have given results for $f_\pi$ and  $m_\pi$ for large quark masses in the curves in Fig.~\ref{fig:5} and the table (Tab.~\ref{tab:largemc}) which can be used to extrapolate to the physical regime.
The NJL model can help to bridge the gap between lattice simulations with heavy chiral fermions and the physical regime.

\acknowledgments

We would like to thank J. Braun for many useful discussions. 
This work is supported in part by the Helmholtz association under grant no. VH-VI-041, in
part by the EU Integrated Infrastructure Initiative Hadron Physics
(I3HP) under contract RII3-CT-2004-506078 and funded in part by  
the German Research Foundation (DFG) under grant no. AL 279/5-1.


\appendix

\section{Flow equation for the pion decay constant}
\label{app:flowsigma0}
In the appendix, we have collected some technical details of the
derivation of the various flow equations used in the
approximate one-loop calculations. 
In this first section, we give the details for the derivation
of the flow equation for the pion decay constant.
We start by considering the minimum condition eq.~\eqref{eq:mincon}
and take the derivative with respect to the renormalization
scale. Note that $\delta$ is independent of the RG scale:
\be
k \frac{\partial}{\partial k}( U_0^\prime(\sigma_0^2,k) - \delta ) \equiv 0 \nn
\\
 k \frac{\partial}{\partial k} U_0' \bigg|_{\Phi^2=\sigma_0^2} \, 2
\sigma_0 + U_0^{\prime \prime} (\sigma_0^2)\,\left(k \frac{\partial}{\partial
  k}\sigma_0^2 \right)  \, 2\, \sigma_0 + U_0^\prime(\sigma_0^2)\, 2 \, \left(k
\frac{\partial}{\partial k} \sigma_0\right)  \equiv 0 \nn\\
 k \frac{\partial}{\partial k} U_0'\bigg|_{\Phi^2=\sigma_0^2} \, 2
\sigma_0 + \frac{4\; \sigma_0^2\, U_0^{\prime \prime} (\sigma_0^2) + 2\,
U_0^\prime(\sigma_0^2)}{2 \sigma_0}  \, \left(k
\frac{\partial}{\partial k} \sigma_0^2 \right)  \equiv 0. 
\label{eq:minconflow}
\ee 
We can identify the term multiplying the derivative of $\sigma_0$ as
the mass of the sigma meson, evaluated at the expectation value $\sigma_0$,
\be
Z_\Phi M^2_\sigma(\sigma_0^2) = 2 U_0^\prime(\sigma_0^2) + 4 \sigma_0^2
U_0^{\prime \prime}(\sigma_0^2).
\ee
In this way, we can express the evolution of the expectation value in
terms of the evolution of the derivative of the potential $U_0(\Phi^2)$
as
\be
k \frac{\partial}{\partial k} \sigma_0^2 &=& -\frac{4 \sigma_0^2}{Z_{\Phi}
  M_\sigma^2(\sigma_0^2)} k\frac{\partial }{\partial k}
  U_0^\prime(\Phi^2) \bigg|_{\Phi^2=\sigma_0^2}. 
\ee
To get the desired differential equation for $U_0'(\Phi^2)$, we
differentiate the flow equation for the potential \eqref{eq:Uk2} and
insert the derivatives of the masses from equations (\ref{eq:MQ})-(\ref{eq:MPI}):
\be
&& k\frac{\partial}{\partial k} U_0^{\prime}(\Phi^2)=\frac{\partial}{\partial \Phi^2}\left( k \frac{\partial}{\partial k}
U_0(\Phi^2)\right) \nn\\
&&= \frac{k^6}{32 \pi^2} \left \{ \frac{4 N_c N_f}{(k^2
 +M_q^2)^2} \frac{\partial M_q^2}{\partial \Phi^2} - \frac{3}{(k^2+
 M_\pi^2)^2} \frac{\partial M_\pi^2}{\partial \Phi^2} - \frac{1}{(k^2+
 M_\sigma^2)^2} \frac{\partial M_\sigma^2}{\partial \Phi^2}
\right\}\nn\\
&&=\frac{k^6}{32
  \pi^2} \left \{ \frac{4 N_c N_f}{(k^2 +M_q^2)^2} \frac{G^2}{Z_q^2} - \frac{3}{2}
\frac{1}{\Phi^2} (M_\sigma^2-M_\pi^2) \left(\frac{1}{(k^2+M_\pi^2)^2} +
  \frac{1}{(k^2+M_\sigma^2)^2} \right)  \right \}.
\ee
For the flow equation for the expectation value of the field
$\Phi$, we evaluate this equation at $\Phi^2=\sigma_0^2$.
In order to obtain a flow equation for the pion decay constant, we now
need to include the flow equation for the coupling $Z_\Phi$:
\be
k \frac{\partial}{\partial k} f_\pi^2 &=& k \frac{\partial}{\partial
  k} (Z_\Phi \sigma_0^2) = \left( k \frac{\partial}{\partial k} Z_\Phi
\right) \sigma_0^2 + Z_\Phi \left( k \frac{\partial}{\partial k}
\sigma_0^2 \right) \nn\\
&=& \frac{k^6}{16 \pi^2} \left\{ - \frac{4 N_c N_f}{(k^2 +M_q^2)^2}
\frac{G^2}{Z_q^2} \sigma_0^2 \left(\frac{1}{(k^2 +M_q^2)} + \frac{2}{M_\sigma^2}
\right) \right . \nn \\
&& \;\;\;\; \left .+ \left(3 \frac{M_\sigma^2- M_\pi^2}{M_\sigma^2} -1
\right)\left(\frac{1}{(k^2 + M_\pi^2)^2} + \frac{1}{(k^2 + M_\sigma^2)^2}
\right)\right .\\
&& \;\;\;\; \left . + \frac{2}{(k^2 + M_\pi^2) (k^2 + M_\sigma^2)}
\right \} \nn
\ee

\section{The chiral condensate $\left\langle\bar{q}q\right\rangle$}
\label{app:qq}

In principle it should be possible to derive the quark mass expansion of
the chiral quark condensate $\left<\bar{q}q\right>$ analytically in a manner analogous to the pion decay constant
$f_\pi$ and the pion mass $m_\pi^2$ in section
\ref{sec:analytical}. For this purpose one has to compute the chiral
condensate from the partition function $Z$,
\begin{align} 
Z[J,\bar{\eta},\eta] =  e^{-W[J,\bar{\eta},\eta]} 
= \int \mathcal{D}q
\mathcal{D}\bar{q} \mathcal{D}\Phi \exp \left( -S[\bar{q},q,\Phi]+
\int d^4 x (J \Phi +\bar{\eta} q + \bar{q} \eta) \right),\label{Zlsm}
\end{align}
by differentiating it with respect to the additional source term $m_c \bar{q}q$
\begin{align} 
\left<\bar{q}q\right> = \frac{\partial}{\partial m_c} \log 
Z(m_c)=\frac{\partial}{\partial m_c} U(\Phi_0^2, m_c).\label{qqfromZ}
\end{align} 
Using renormalization group methods, this would lead to a flow equation
for $\left<\bar{q}q\right>$ similar to equations \eqref{fpik} and \eqref{mpik}
for $f_\pi$ and $m_\pi^2$.
Unfortunately, it turns out that we probably miss some relevant terms
in equation \eqref{qqfromZ}. This may be related to a lack of
rebosonization of the evolving four-fermion term \cite{Gies:2001nw}. Therefore we
restrict ourself to show that the $\left<\bar{q}q\right>$-expansion
follows from the other two expansions.
We derive $\left<\bar{q}q\right>$ from our results for $f_\pi$ and $m_\pi^2$ from
section \ref{sec:analytical} by using the Gell-Mann--Oakes--Renner
relation, $m^2_\pi=\frac{\left<\bar{q}q \right>}{f_\pi^2}m_c$, and then
replace $m_c$ by the lowest-order relation in the chiral expansion,
$m_c=M^2 \frac{f_{\pi 0}^2}{\left<\bar{q}q \right>_0}$. This gives
\begin{align}
\left<\bar{q}q \right>&=\frac{m^2_\pi f_\pi^2}{m_c}=\frac{m^2_\pi}{M^2} \frac{f_\pi^2}{f_{\pi 0}^2} \left<\bar{q}q \right>_0\nn\\
&=\left<\bar{q}q
\right>_0 \left(1-\frac{3 M^2}{32 \pi^2 f_{\pi 0}^2} \log \frac{M^2}{k_{\chi SB}}+{\cal  O}(M^2)\right).
\end{align}
As for the pion decay constant and the pion mass, we reproduce exactly the
logarithmic term of $\chi$PT (cf. eq.(\ref{qq})). In the way employed here,
this is a consistency check of our expansions for
$f_\pi$ and $m_\pi^2$ (eqs. \eqref{f-pientw} and \eqref{m-pientw})
with the Gell-Mann--Oakes--Renner relation.

\section{Regularization independence of the chiral logarithms}

In general, for finite values of the renormalization scale $k$, the
results for the RG flow depend on the choice of regularization
function. However, for the family of cutoff functions that we consider
in this paper, 
\be
f^{(a)}(\tau k^2) &=& \left(\sum_{j=0}^{a} \frac{1}{j!} (\tau
k^2)^j\tau  \right) \exp(-\tau k^2), \label{eq:cutoffa}
\ee
we can show that the logarithmic contributions
involving the pion mass are independent of the choice of function
within this family.
The general flow equation for the potential, which results from a
cutoff function of this type for $a \ge 2$, is
\be
k \partial_k U_k(\Phi^2)&=& \frac{(k^2)^{a+1}}{16 \pi^2 a (a-1)}\left[
  \frac{1}{(k^2 + M_\sigma^2(\Phi^2))^{a-1}} + \frac{3}{(k^2 +
    M_\pi^2(\Phi^2))^{a-1}} - \frac{4 N_c N_f} {(k^2 +
    M_q^2(\Phi^2))^{a-1}} \right]. \nn\\
\ee
As an example, consider the pion contributions to the flow equation
for the expectation value of the field, $\sigma_0(k)$:
\be
[k \partial_k \sigma_0(k)]_{\mbox{\scriptsize pion}} &=& \frac{2
  \lambda \sigma_0(k)}{M_\sigma^2(\sigma_0(k)^2)} \frac{3}{16 \pi^2} k^2
\frac{1}{a} \frac{(k^2)^a}{(k^2+ M_\pi^2(\sigma_0(k)^2))^{a}}.    
\ee
For simplicity we neglect the couplings beyond the four-point
couplings as in equation \eqref{U0par} and approximate $\frac{\partial M_\pi^2}{\partial \Phi^2} =
\lambda$.

The term relevant for the logarithmic contribution is
\be
\frac{1}{a} \frac{(k^2)^a}{(k^2+ M_\pi^2)^{a}} &=& \frac{1}{a}
\frac{(k^2+ M_\pi^2 -M_\pi^2)^a}{(k^2+ M_\pi^2)^{a}} \nonumber \\
&=& \frac{1}{a} \frac{(k^2 + M_\pi^2)^a -
    \left({a \atop 1} \right) M_\pi^2 (k^2
    +M_\pi^2)^{a-1} + \left({ a \atop 2} \right)
    (M_\pi^2)^2 (k^2 +M_\pi^2)^{a-2} +- \cdots  }{(k^2+ M_\pi^2)^{a}}
     \nonumber \\
&=& \frac{1}{a} - M_\pi^2 \frac{1}{(k^2 + M_\pi^2)} + \frac{(a-1)}{2}
      (M_\pi^2)^2 \frac{1}{(k^2 + M_\pi^2)^2} -+ \cdots  
\ee 
We note that the second term is independent of the parameter
$a$. After setting $M_\pi^2= M^2$ and on
integration over the scale $k$, this term yields
\be
- \frac{1}{2} M^2 \log(k^2 + M^2), 
\ee
and is thus responsible for the emergence of the chiral log. 
We conclude that in the current approach, the logarithmic
contribution is independent of the
regularization scheme. 

In general, the expansion of the observables in terms of the pion
mass parameter $M^2$ is dependent on the choice of the
cutoff function in our perturbative approximation. However, the
leading logarithmic term in the pion mass parameter (the chiral log)
is independent of the choice of $a$ for the cutoff function.
 
 \section{Results for large quark masses}
 \label{app:table}
In the following, we list selected results for $f_\pi$ and $m_\pi$ from the RG calculations for large values of the current quark mass. We give results for several different UV cutoff values. 
 
\begin{table}[t]
\begin{tabular}{||c|cc|cc|cc||}
\hline
\hline
 & $\Lambda_{\scriptstyle{UV}}=1000$ & ($m_{\mathrm{phys}}=12.9$)&$\Lambda_{\scriptstyle{UV}}=1200$ &($m_{\mathrm{phys}}=9.0$) &$\Lambda_{\scriptstyle{UV}}=1500$ & ($m_{\mathrm{phys}}=6.0$) \\
\hline
$m_c$ &  $f_\pi$& $m_\pi$& $f_\pi$&$m_\pi$& $f_\pi$& $m_\pi$\\
\hline
10& $\phantom{1}91.4$&$120.9$ &$\phantom{1} 93.1$ &$144.8$ & $\phantom{1} 97.9$&$175.9$\\
20& $\phantom{1} 94.7$&$172.3$ & $\phantom{1} 99.3$&$204.3$ & $108.0$&$244.6$\\
40& $\phantom{1} 99.4$&$247.1$ & $107.7$&$289.1$ & $120.9$&$341.1$\\
60& $102.9$&$306.9$ & $113.5$&$355.9$ & $129.7$&$416.2$\\
80& $105.5$&$359.6$ & $117.9$&$414.2$ & $136.3$&$481.0$\\
100& $107.5$&$408.1$ & $121.3$&$467.4$ & $141.4$&$539.7$\\
120& $109.1$&$454.2$ & $124.0$&$517.4$ & $145.7$&$594.4$\\
\hline
\hline
\end{tabular}
\caption{\label{tab:largemc} Values for the pion decay constant
   $f_\pi$ and the pion mass $m_\pi$ for large current quark masses
   $m_c$ from the
   RG calculation. The data are given for three different UV cutoff choices
   $\Lambda_{\scriptstyle{UV}}$. All values are given in [MeV].}
\end{table}
\newpage
\bibliography{rg,chpt}

\end{document}